\documentclass[aps,preprintnumbers,amsmath,amssymb,nofootinbib,eqsecnum, preprintnumbers]{revtex4}
\usepackage{eurosym}
\usepackage{amsfonts}
\usepackage{amsmath}
\usepackage{amssymb,epsf}
\usepackage{color}
\usepackage{graphicx}
\usepackage{natbib}
\usepackage{float}
\usepackage{caption}
\usepackage{subfig}
\usepackage{epstopdf}

\begin{document}

 \title{Black holes in dRGT massive gravity with the signature of EHT observations of M87*}

 \author{
 S. H. Hendi$^{1,2,3}$\footnote{email address: hendi@shirazu.ac.ir},
 Kh. Jafarzade$^{1,2,5}$\footnote{email address: khadije.jafarzade@gmail.com} and
 B. Eslam Panah$^{4,5,6}$\footnote{email address: eslampanah@umz.ac.ir}}

\affiliation{
 $^{1}$Department of Physics, School of Science,Shiraz University, Shiraz 71454, Iran\\
 $^{2}$Biruni Observatory, School of Science, Shiraz University, Shiraz 71454, Iran\\
 $^{3}$Canadian Quantum Research Center 204-3002 32 Ave Vernon, BC V1T 2L7 Canada\\
 $^4$Department of Theoretical Physics, Faculty of Science, University of Mazandaran, P. O. Box 47416-95447, Babolsar, Iran\\
 $^5$ICRANet-Mazandaran, University of Mazandaran, P. O. Box 47416-95447, Babolsar, Iran\\
 $^6$ICRANet, Piazza della Repubblica 10, I-65122 Pescara, Italy}

\begin{abstract}
The recent Event Horizon Telescope (EHT) observations of the M87*
have  led to a surge of interest in studying the shadow of black
holes. Besides, investigation of time evolution and lifetime of
black holes helps us to veto/restrict some theoretical models in
gravitating systems. Motivated by such exciting properties, we
study optical features of black holes, such as the shadow
geometrical shape and the energy emission rate in modified
gravity. We consider a charged AdS black hole in dRGT massive
gravity and look for criteria to restrict the free parameters of
the theory. The main goal of this paper is to compare the shadow
of the mentioned black hole in a rotating case with the EHT data
to obtain the allowed regions of the model parameters. Therefore,
we employ the Newman-Janis algorithm to build the rotating
counterpart of static solution in dRGT massive gravity. We also
calculate the energy emission rate for the rotating case and
discuss how the rotation factor and other parameters affect the
emission of particles around the black holes.
\end{abstract}

\maketitle

\section{Introduction}

The existence of a mysterious object known as a black hole (BH) is
one of the important predictions of General Relativity (GR). In
this regard, it has been proved that a BH really can form using
ingenious mathematical methods by Penrose in 1965 \cite{Penrose}.
One of the interesting records of the existence of a BH is related
to the gravitational wave, which was detected by the advanced
LIGO/Virgo collaboration in 2016 \cite{LVCI}. This record provides
a constraint on the modified theories of gravity, for example, the
existence of a tight bound on the graviton mass \cite{LVCII}, or
the speed of gravitational wave \cite{Cornish}. Another strong
record is related to the first image of a BH, which was detected
by the Event Horizon Telescope in 2019 \cite{Akiyama}. It may
impose some constraint on properties of BHs
\cite{constrainETHI,constrainETHII,constrainETHIII,constrainETHIV,constrainETHV},
and parameters of modified theories of gravity
\cite{ModM87I,ModM87II}. These reasons motivate us to study
theoretical BHs in the light of observational data, which are
receiving much attention today.

Some interesting modified theories of gravity have been proposed
since our Universe expands with an acceleration
\cite{AccUniI,AccUniII}, whereas GR cannot describe it, properly.
Massive gravity is one of the extraordinary theories of gravity,
which modifies gravitational effects by weakening it at a large
scale compared to GR. On the other hand, to satisfy the concrete
observations, it reduces to GR at small scales. The nonzero mass
of gravitons in massive gravity has an upper limit obtained by
recent observations of the advanced LIGO/Virgo collaboration
\cite{LVCII}. In addition, it was indicated that the mass of
graviton might be variable. Indeed, the mass of gravity is tiny in
the usual weak gravity environments. In contrast, it becomes much
more significant in the strong gravity regime such as compact
objects and BHs \cite{Variablegraviton}. Appropriate description
of rotation curves of the Milky Way and spiral galaxies
\cite{Panpanich}; the existence of massive neutron stars
\cite{neutronS} and super-Chandrasekhar white dwarfs
\cite{whitedwarf}; explaining the current observations related to
dark matter \cite{darkMI,darkMII}; the existence of a remnant for
BH \cite{remnantI,remnantII}, are some interesting achievements in
the context of massive gravity.

In 1939, Fierz and Pauli introduced a class of massive gravity
theory in the flat background, in which the interaction terms
added to GR \cite{FPmassive}. This theory of massive gravity
suffers from a discontinuity, which is known as van
Dam-Veltman-Zakharov (vDVZ) discontinuity \cite{vDVZI,vDVZII}. In
order to remove this discontinuity, nonlinear-massive gravity was
introduced by Vainshtein \cite{Vainshtein}. Afterward, Boulware
and Deser have shown that such nonlinear generalizations generate
a ghost instability, which is known as Boulware-Deser (BD) ghost
\cite{BDghost}. Arkani-Hamed et al. have resolved these problems
by introducing St\"{u}ckelberg fields, which lead to a class of
potential energies depending on the gravitational metric and an
internal Minkowski metric (reference metric) \cite{Arkani-Hamed}.
Finally, a new version of massive gravity was introduced by de
Rham, Gabadadze, and Tolley (dRGT) \cite{dRGTI,dRGTII}, which is
free of vDVZ discontinuity and BD ghost in arbitrary dimensions
\cite{HassanI}. This theory of massive gravity is known as dGRT
massive gravity. Notably, there are two metric tensors in dRGT
massive gravity, one of which is a fixed spacetime background. The
generalization of fixed metric to dynamical metric was expanded by
Hassan and Rosen, later known as the bimetric or bi-gravity theory
\cite{bimetric}. However, we consider the original massive gravity
with a fixed background (dRGT massive gravity) in this work.

The establishment of a firm theoretical foundation of dRGT massive
gravity has opened the possibility of studying its astrophysical
and cosmological applications. In this regard, BH solutions with
interesting properties have been obtained by considering the
massive gravity in Refs.
\cite{BHmassI,BHmassVI,BHmassIX,BHmassXII,BHmassXIV}.
In the astrophysics context, the properties of compact objects
such as relativistic stars \cite{massstarI,massstarII}, neutron
stars \cite{neutronS,neutronSI}, and white dwarfs
\cite{whitedwarf}, in this massive gravity have been evaluated.
Considering dRGT massive gravity, the effect of helicity-0 mode,
which remains elusive after analysis of cosmological perturbation
around an open Friedmann--Lemaitre--Robertson--Walker universe,
was evaluated in Ref. \cite{Chiang}. In addition, the cosmological
behavior \cite{Leon}, bounce and cyclic cosmology \cite{Cai},
graviton mass as a candidate for dark matter \cite{Aoki}, and
other studies have been done for finding some constraints on
parameters of massive gravity by considering observational
cosmological data
\cite{consII,consVI}.

Recently, research in astrophysical BHs has gained much attention
due to the breakthrough discovery of the reconstruction of the
event-horizon-scale images of the supermassive BH in the galaxy
M87* by the EHT project \cite{Akiyama,Akiyama2019b}. It was the
first direct evidence of the existence of BHs compatible with the
prediction of GR. Although the first image of the BH does not
allow one to identify the BH geometry clearly, the principal
strategy for improving measurements should lead to much higher
resolution in the future \cite{Goddi2016}. So, one cannot
underestimate theoretical efforts to calculate forms of shadows
cast by BHs in various theories of gravity and astrophysical
environments. The BH shadow is the optical appearance feature of
the BH when a bright distant source is behind it. It appears as a
two-dimensional dark zone for a distant observer, like us on
Earth. The shadow of a BH can provide us with information about
the BH and serve as a useful tool for testing GR. For a
non-rotating BH, the boundary of the shadow is a perfect circle
and was first studied by Synge \cite{Synge1966} and later by
Luminet, who further considered the effect of a thin accretion
disk on the shadow \cite{Luminet1979}. While the rotating case has
an elongated shape in the direction of the rotation axis due to
the dragging effect \cite{Bardeen1973}. The BH shadow has become a
hot topic among researchers for the simple fact to evaluate the
soon-expected observational data best. Shadows in various BH
spacetimes have been studied extensively in the literature in the
last decades \cite{Tsukamoto2018,Wei2013,Bambi2010}. This topic
has also been extended to BHs in modified GR
\cite{Amarilla2010,Kumar2021}, to BHs with higher or extra
dimensions \cite{Papnoi2014,Pratap2017}, and BHs surrounded by
plasma \cite{Atamurotov2015}. After the EHT announcement, a large
amount of research has been devoted to calculating the shadows of
a vast class of BH solutions and the confrontation with the
extracted information from the EHT BH shadow image of M87*
\cite{Davoudiasl2019,Konoplya2019}.

 Although, in preliminary studies in the context of
black hole shadow, the BH was assumed to be eternal, i.e.,
spacetime was assumed to be time-independent, recent astronomical
observations have indicated that our Universe is currently
undergoing a phase of accelerated expansion. This reveals the fact
that the BH shadow can be dependent on time. Although the effect
of the cosmological expansion is negligible for the BH candidates
at the center of the Milky Way galaxy and at the centers of nearby
galaxies,  its influence on the shadow size will be significant
for galaxies at a larger distance \cite{Perlick2018}. In the
context of standard cosmology, based on GR, this acceleration
cannot be explained unless by introducing an unknown energy
component so-called \emph{dark energy}. Another way to explain
such acceleration is the modification of GR. One of the well-known
candidates for dark energy scenarios is the cosmological constant.
Recently, the role of the cosmological constant in gravitational
lensing has been extensively studied
\cite{Piattella2016,Butcher2016,Faraoni2017,Zhang2017bc,Zhao2015ab}.
As we know, the shadow and photon ring are caused by the light
deflection or gravitational lensing by BHs. So, one can inspect
the effect of the cosmological constant on the shadow of black
holes \cite{Haroon2019,Firouzjaee2019}. It is worth noting that
although the expansion of the Universe was based on a positive
cosmological constant, there is some evidence showing that it can
be associated with a negative cosmological constant. The first one
is through observational Hubble constant data. As we know, an
interesting method for investigating the accelerated cosmic
expansion and studying dark energy properties is through the
Hubble constant, measured as a function of cosmological redshift
\cite{Zhang2007fd,Wan2007mn,Ma2011kj}. Investigating the behavior
of $ H(z) $ in low redshift data showed that the dark energy
density has a negative minimum for certain redshift ranges, which
can be simply modeled via a negative cosmological constant
\cite{Dutta2020}. The second reason is through supernova data.
Although according to the high-redshift supernova, the expansion
of the Universe is accelerating due to a positive cosmological
constant, it should be noted that the supernova data themselves
derive a negative mass density in the Universe
\cite{AccUniI,AccUniII} which can be equivalent to a negative
cosmological constant \cite{Farnes2018}. Several galaxy cluster
observations have inferred the presence of a negative mass in
cluster environments. According to Ref. \cite{Farnes2018},  the
introduction of negative masses can lead to an AdS space. Another
reason to consider a negative cosmological constant is the concept
of stability of the accelerating Universe. The possibility of de
Sitter expanding spacetime with a constant internal space was
investigated in Ref. \cite{Maeda2014a} and was shown that the de
Sitter solution would be stable just in the presence of the
negative cosmological constant.

The manuscript is structured as follows: In Sec. \ref{Sec. II}, we
briefly review the charged AdS BH solution in dRGT massive
gravity. In Subsec. \ref{Sec. IIA}, we determine the null
geodesics equations as well as the radius of the photon sphere and
shadow, and we investigate the ratio of shadow radius and photon
sphere to find an acceptable optical behavior. In Subsec.
\ref{Sec. IIB}, we calculate the energy emission rate and explore
the effect of different parameters on the emission of particles
around the BH. In Sec. \ref{Sec. III}, by applying the
Newman-Janis algorithm, we obtain the rotating charged AdS BH
solution in dRGT massive gravity. The shadow geometrical shape and
the energy emission rate for this type of solution are
investigated in Subsec. \ref{Sec. IIIA} and Subsec. \ref{Sec.
IIIB}, respectively. In Sec. \ref{Sec. IV}, we confront the
obtained shadow of the rotating charged case with the EHT image of
the supermassive object in M87*, and we extract explicit
constraints on the parameters of the dRGT massive theory. We
eventually summarize our results and conclude in Sec. \ref{Sec.
V}.

 \section{Charged BHs in dRGT massive gravity} \label{Sec. II}

The action of dRGT massive gravity is expressed as a Hilbert-Einstein action
plus suitable nonlinear interaction terms interpreted as a graviton mass,
which is given by \cite{dRGTII}
\begin{equation}
I=\frac{1}{16\pi }\int d^{4}x\sqrt{-g}(R-\mathcal{F}+m_{g}^{2}\mathcal{U}(g,\phi ^{a})),
\label{action:new}
\end{equation}%
where $ \mathcal{F}\equiv F_{\mu\nu} F^{\mu\nu}$ , and $
F_{\mu\nu} $ is the Faraday tensor which is constructed using the
$ U(1) $ gauge field $ A_{\nu} $ as $ F_{\mu\nu} =\partial
_{[_{\mu}A_{\nu}]}$. Also $R$ and $\mathcal{U}$ are, respectively,
the Ricci scalar and the effective potential of the graviton,
which modifies the gravitational sector with a graviton mass
$m_{g}$. The potential $\mathcal{U}$ in four-dimensional spacetime
is of the form
\begin{equation*}
\mathcal{U}(g,\phi ^{a})=\mathcal{U}_{2}+\alpha _{3}\mathcal{U}_{3}+\alpha
_{4}\mathcal{U}_{4},
\end{equation*}%
in which $\alpha _{3}$ and $\alpha _{4}$ are dimensionless free parameters
of the theory. The functional form of $\mathcal{U}_{i}$ with respect to the
metric $g$ and scalar field $\phi ^{a}$ can be expressed as

\begin{eqnarray}
\mathcal{U}_{2} &\equiv &[\mathcal{K}]^{2}-[\mathcal{K}^{2}],  \notag \\
&&  \notag \\
\mathcal{U}_{3} &\equiv &[\mathcal{K}]^{3}-3[\mathcal{K}][\mathcal{K}^{2}]+2[%
\mathcal{K}^{3}],  \notag \\
&&  \notag \\
\mathcal{U}_{4} &\equiv &[\mathcal{K}]^{4}-6[\mathcal{K}]^{2}[\mathcal{K}%
^{2}]+8[\mathcal{K}][\mathcal{K}^{3}]+3[\mathcal{K}^{2}]^{2}-6[\mathcal{K}%
^{4}],
\end{eqnarray}%
in which
\begin{equation*}
\mathcal{K}_{\,\,\,\nu }^{\mu }=\delta _{\nu }^{\mu }-\sqrt{g^{\mu \sigma
}f_{ab}\partial _{\sigma }\phi ^{a}\partial _{\nu }\phi ^{b}},
\end{equation*}%
where $f_{ab}$ is an appropriate non-dynamical reference metric and the
rectangular bracket denotes the traces, namely $[\mathcal{K}]=\mathcal{K}%
_{\,\,\,\mu }^{\mu }$, and $[\mathcal{K}^{n}]=(\mathcal{K}^{n})_{\,\,\,\mu
}^{\mu }$.  The scalar fields $\phi ^{a}$ (called St\"{u}ckelberg
fields), are introduced in order to restore the general covariance of the
theory. The corresponding theory described by action (\ref{action:new}) propagates 7 degrees of freedom (see Appendix A for more detail).

The gravitational and electromagnetic filed equations of the theory are obtained as
\begin{eqnarray}
G_{\mu \nu }+m_{g}^{2}X_{\mu \nu }&=&-\frac{1}{2}g_{\mu\nu}\mathcal{F}+2F_{\mu\lambda}F_{\nu}^{\lambda},\\
 \nabla_{\nu}F^{\mu\nu}&=&0,
 \label{EqEFE}
\end{eqnarray}%
where $G_{\mu \nu }$ is the Einstein tensor and the tensor $X_{\mu \nu }$
can be interpreted as the effective energy-momentum tensor in the following
form
\begin{equation}
X_{\mu \nu }=\mathcal{K}_{\mu \nu }-\mathcal{K}g_{\mu \nu }-\alpha \left(
\mathcal{K}_{\mu \nu }^{2}-\mathcal{K}\mathcal{K}_{\mu \nu }+\frac{\mathcal{U%
}_{2}}{2}g_{\mu \nu }\right) +3\beta \left( \mathcal{K}_{\mu \nu }^{3}-%
\mathcal{K}\mathcal{K}_{\mu \nu }^{2}+\frac{\mathcal{U}_{2}}{2}\mathcal{K}%
_{\mu \nu }-\frac{\mathcal{U}_{3}}{6}g_{\mu \nu }\right) ,  \label{effemt}
\end{equation}%
where two parameters $\alpha _{3}$ and $\alpha _{4}$ are
reparameterized by introducing two new parameters $\alpha $ and
$\beta $ as follows
\begin{equation}
\alpha _{3}=\frac{\alpha -1}{3}~,~~~\&~~~\alpha _{4}=\frac{\beta }{4}+\frac{%
1-\alpha }{12},  \label{alphabeta}
\end{equation}%
in which $\alpha $ and $\beta $ are two arbitrary dimensionless constants.

We consider a static and spherical symmetric spacetime in $4$-dimensional as
\begin{equation}
\text{d}s^{2}=-f(r)\text{d}t^{2}+f^{-1}(r)\text{d}r^{2}+r^{2}\text{d}\Omega
^{2},  \label{Metric}
\end{equation}%
where d$\Omega ^{2}=$d$\theta ^{2}+\sin ^{2}\theta $d$\varphi ^{2}$.

In order to obtain exact solutions, we consider a singular reference metric in the following form
\begin{equation}
f_{\mu \nu }=diag\left( 0,0,h^{2},h^{2}\sin ^{2}\theta \right) ,
\label{refrenceM}
\end{equation}
where $h$ is a positive constant with the dimension of length. Since the reference metric (\ref{refrenceM}) depends only on the spatial components, general covariance is preserved in the $ t $ and $ r $ coordinates, but is broken in the two spatial dimensions.
One can also imagine a more general reference metric that does not
respect diffeomorphism invariance in the $r-$direction. For
instance, to preserve rotational invariance on the sphere and
general time reparametrization invariance, a natural ansatz can be
$f_{\mu\nu} = diag(0, 1, h^{2}, h^{2}sin^{2}\theta)$. One might
also break diffeomorphism invariance in the $r-$direction by
considering a different generalization of $ f_{\mu\nu} $, with
$sin^{2}\theta f_{\theta \theta} = f_{\varphi \varphi}= F(r)$, and
all other components to be zero \cite{Adams2015}. This can lead to
an ability to add arbitrary polynomial terms in $r$ to the
emblackening factor. From what was expressed, one can find that
for each choice of the reference metric one is essentially dealing
with a different theory. In other words, considering different possibilities for the reference metric leads to a variety of new
solutions. From a gauge/gravity
duality perspective, massive gravity on AdS with a singular
(degenerate) reference metric is dual to homogenous and isotropic
condensed matter systems which leads to a boundary theory with the
finite direct-current (DC) conductivity
\cite{Blake2013,Vegh2013,Davison2013}, the desired property for
normal conductors that is absent in massless gravities
\cite{Hartnoll2008,Gregory2009}. So, massive gravity
with this choice of reference metric can be of particular interest to researchers. Although such an investigation is not the purpose of this paper.

Using the field equation (\ref{EqEFE}), and metrics (\ref{Metric}) and (\ref%
{refrenceM}), the BH solution is obtained as \cite{BHmassX}
\begin{equation}
f(r)=1-\frac{2M}{r}+\frac{Q^{2}}{r^{2}}-\frac{\Lambda }{3}r^{2}+\gamma
r+\varepsilon ,  \label{Eqsol}
\end{equation}%
where $M$ and $Q$ are, respectively, the mass and electric charge
of the BH and other parameters are defined as follows
\begin{eqnarray}
-\Lambda  &=&3m_{g}^{2}(1+\alpha +\beta ),  \notag \\
&&  \notag \\
\gamma  &=&-hm_{g}^{2}(1+2\alpha +3\beta ),  \notag \\
&&  \notag \\
\varepsilon  &=&h^{2}m_{g}^{2}(\alpha +3\beta ).
\end{eqnarray}

This solution contains various signatures of other well-known BH
solutions found in the literature. By setting $m_{g}=0$, we have
the Reissner-Nordstr\"{u}m solution. For $h=0$, which sets $\gamma
=\varepsilon =0$, the solution can be classified according to the
values of $\alpha $ and
$\beta $. For $1+\alpha +\beta <0$, the solution becomes the Reissner-Nordstr%
\"{u}m-dS solution, while the case $1+\alpha +\beta >0$ yields the
Reissner-Nordstr\"{u}m-AdS solution. The linear term $\gamma r$ is a
characteristic term of this solution, which distinguishes it from other
solutions. The constant potential $\varepsilon $, corresponds to the global
monopole term which naturally emerges from the graviton mass. Usually, a
global monopole solution comes from a topological defect in high energy
physics at early universe resulting from a gauge-symmetry breaking \cite{Huang1a,Tamaki1a}.

\subsection{Photon sphere and shadow}
\label{Sec. IIA} In this section, we would like to analyze the
motion of a free photon in the BH background (\ref{Eqsol}). To do
so, we employ the geodesic equation to calculate the radius of the
innermost circular orbit for a photon in the BH spacetime. The
Hamiltonian of the photon moving in the static spherically
symmetric spacetime is expressed as \cite{Carter1a,Decanini1a}
\begin{equation}
H=\frac{1}{2}g^{ij}p_{i}p_{j}=0.  \label{EqHamiltonian}
\end{equation}

Due to the spherically symmetric property of the BH, we consider
trajectories of photons on the equatorial plane with $\theta =\pi
/2$. Thus, Eq. (\ref{EqHamiltonian}) can be written as
\begin{equation}
\frac{1}{2}\left[ -\frac{p_{t}^{2}}{f(r)}+f(r)p_{r}^{2}+\frac{p_{\varphi }^{2}}{%
r^{2}}\right] =0.  \label{EqNHa}
\end{equation}

Since the Hamiltonian does not depend explicitly on the coordinates $t $ and
$\varphi $, one can define $p_{t} $ and $p_{\varphi} $ as constants of motion. We
consider $p_{t}=-E$ and $p_{\varphi}=L $, where $E $ and $L $ are the energy
and angular momentum of the photon, respectively.

Using the Hamiltonian formalism, the equations of motion are obtained as
\begin{equation}
\dot{t}=\frac{\partial H}{\partial p_{t}}=-\frac{p_{t}}{f(r)},~~~\&~~~\dot{r}%
=\frac{\partial H}{\partial p_{r}}=p_{r}f(r),~~~\&~~~\dot{\varphi}=\frac{%
\partial H}{\partial p_{\varphi }}=\frac{p_{\varphi }}{r^{2}},
\end{equation}%
where the overdot is the derivative with respect to the affine parameter and
$p_{r}$ is the radial momentum. Using the equations of motion and two
conserved quantities, one can rewrite the null geodesic equation as follows
\begin{equation}
\dot{r}^{2}+V_{\mathrm{eff}}(r)=0,  \label{EqVef1}
\end{equation}%
where $V_{\mathrm{eff}}$ is the effective potential of the photon, given by
\begin{equation}
V_{\mathrm{eff}}(r)=f(r)\left[ \frac{L^{2}}{r^{2}}-\frac{E^{2}}{f(r)}\right]
.  \label{Eqpotential}
\end{equation}

For a circular null geodesic, the effective potential satisfies the
following conditions, simultaneously

\begin{equation}
V_{\mathrm{eff}}(r_{ph})=0,~~~\&~~~V_{\mathrm{eff}}^{\prime }(r_{ph})=0,
\label{EqVeff1}
\end{equation}%
where the first condition determines the critical angular momentum of the
photon sphere $L_{p}$, while the second condition yields the photon sphere
radius ($r_{ph}$). It should be noted that the photon orbits are unstable
and are determined by the condition $V_{\mathrm{eff}}^{\prime \prime
}(r_{ph})<0$.

Taking into account the effective potential (\ref{Eqpotential}), $V_{\mathrm{%
eff}}^{\prime }(r_{ph})=0$ leads to the following relation
\begin{equation}
\gamma r_{ph}^{3}+2(1+\varepsilon )r_{ph}^{2}-6Mr_{ph}+4Q^{2}=0,
\label{Eqrph1}
\end{equation}%
as we see, Eq. (\ref{Eqrph1}) is cubic in $r_{ph}$ and its discriminant $%
\Delta $ can be obtained as follows

\begin{figure}[H]
\centering
\subfloat[$ Q=0.2 $]{
        \includegraphics[width=0.31\textwidth]{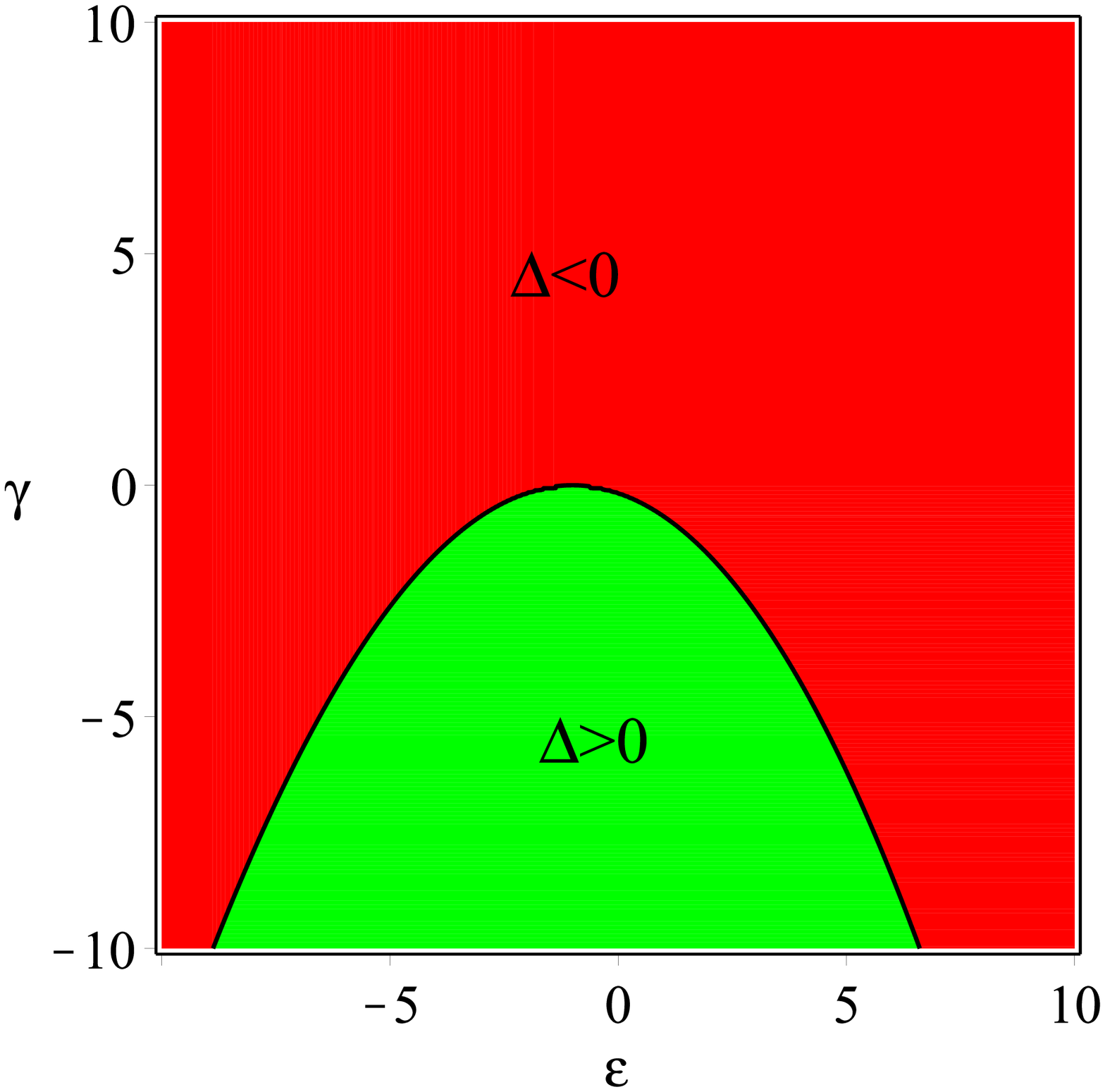}}
\subfloat[$ Q=0.5 $]{
        \includegraphics[width=0.31\textwidth]{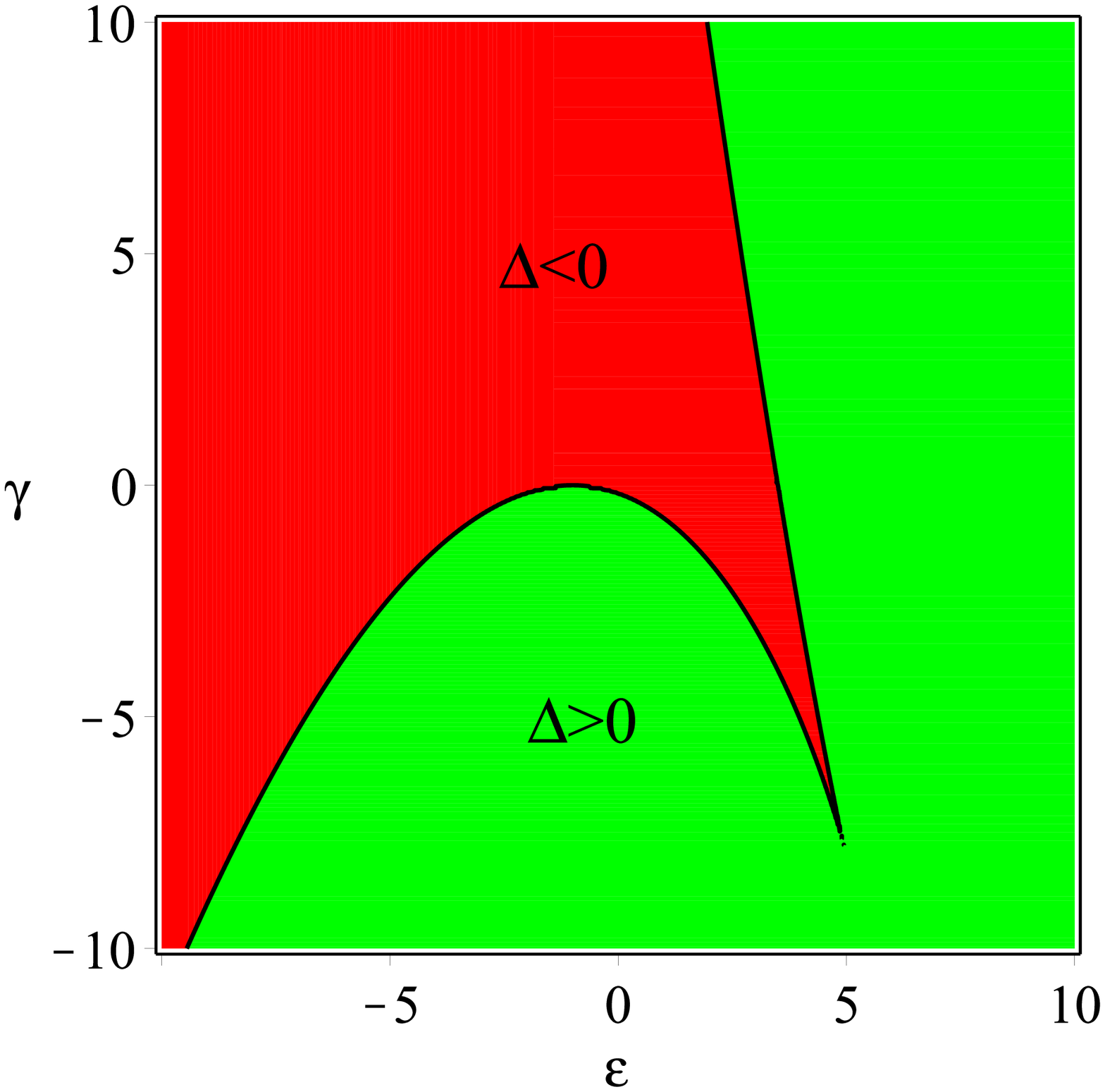}}
\subfloat[$ Q=0.8 $]{
        \includegraphics[width=0.31\textwidth]{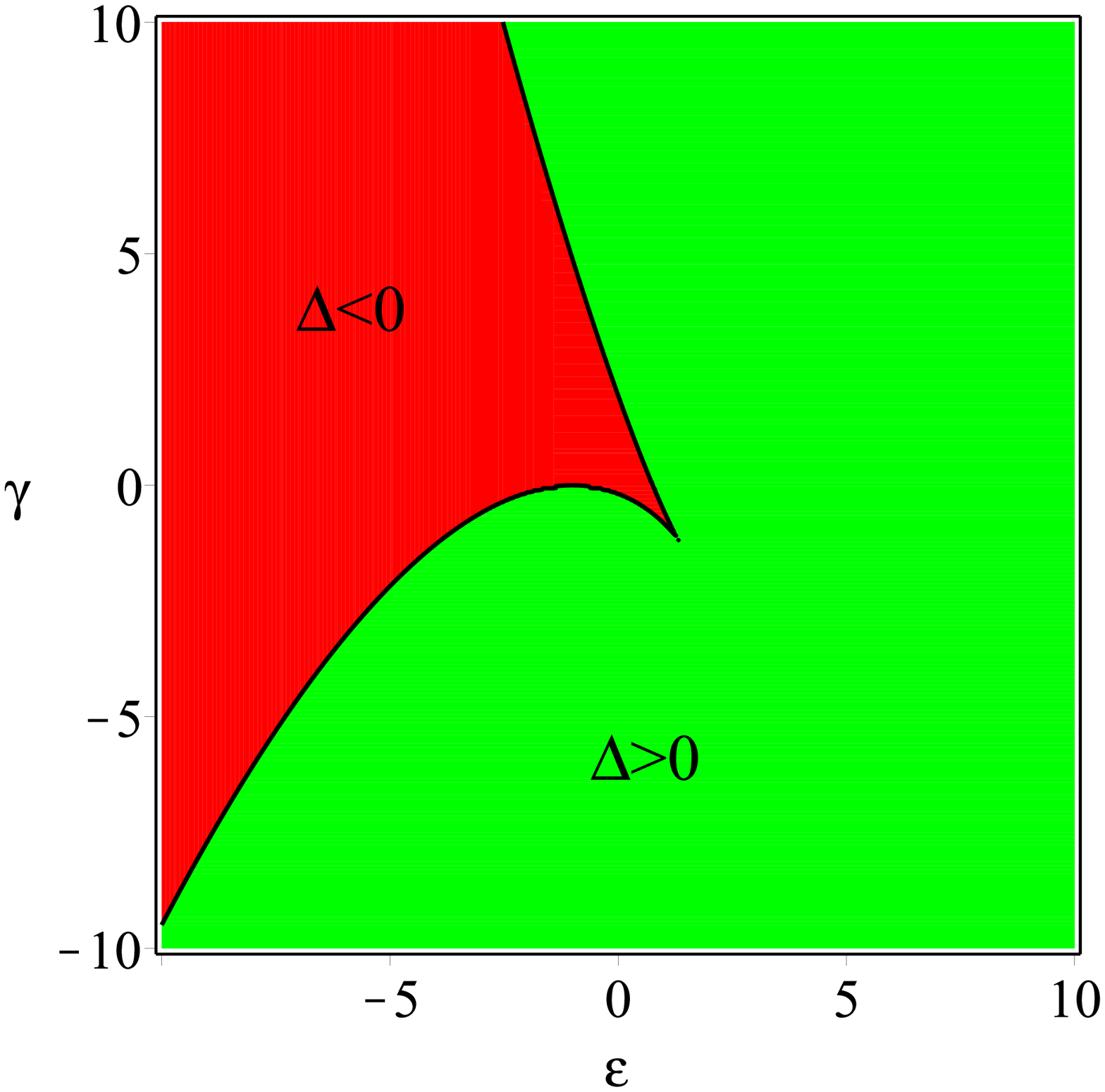}}\newline
\caption{Positive or negative regions of discriminant $\Delta $ for $M=1$.}
\label{Fig1}
\end{figure}
\begin{equation}
\Delta =\frac{p^{3}}{27}+\frac{q^{2}}{4},  \label{EqDEl}
\end{equation}%
with $p$ and $q$ defined by
\begin{eqnarray}
p &=&-\frac{6M}{\gamma }-\frac{4}{3}\frac{(1+\varepsilon )^{2}}{\gamma ^{2}},
\\
&&  \notag \\
q &=&\frac{4Q^{2}}{\gamma }+\frac{4M(1+\varepsilon )}{\gamma ^{2}}+\frac{16}{%
27}\frac{(1+\varepsilon )^{3}}{\gamma ^{3}}.  \label{spq}
\end{eqnarray}

Depending on the values of the BH parameters, $\Delta $ can be
positive or negative. In order to have a more precise picture, we
have distinguished regions with negative/positive $\Delta $ in
Fig. \ref{Fig1}.

If $\Delta >0 $, in this case there is only one real solution given by
\begin{equation}
r_{ph}=\left(-\frac{q}{2}+\sqrt{\Delta} \right)^{\frac{1}{3}}+\left(-\frac{q%
}{2}-\sqrt{\Delta} \right)^{\frac{1}{3}}-\frac{2}{3}\frac{(1+\varepsilon)}{%
\gamma}.  \label{Eqroot1}
\end{equation}

If $\Delta <0$, three real solutions exist in this case which are determined
as follows
\begin{eqnarray}
r_{ph}^{(1)} &=&\frac{2\sqrt{-p}}{\sqrt{3}}\sin \left[ \frac{1}{3}\sin
^{-1}\left( \frac{3\sqrt{3}q}{2(\sqrt{-p})^{3}}\right) \right] -\frac{2}{3}%
\frac{(1+\varepsilon )}{\gamma }, \\
&&  \notag \\
r_{ph}^{(2)} &=&-\frac{2\sqrt{-p}}{\sqrt{3}}\sin \left[ \frac{1}{3}\sin
^{-1}\left( \frac{3\sqrt{3}q}{2(\sqrt{-p})^{3}}\right) +\frac{\pi }{3}\right]
-\frac{2}{3}\frac{(1+\varepsilon )}{\gamma }, \\
&&  \notag \\
r_{ph}^{(3)} &=&\frac{2\sqrt{-p}}{\sqrt{3}}\cos \left[ \frac{1}{3}\sin
^{-1}\left( \frac{3\sqrt{3}q}{2(\sqrt{-p})^{3}}\right) +\frac{\pi }{6}\right]
-\frac{2}{3}\frac{(1+\varepsilon )}{\gamma }.  \label{Eqroot3}
\end{eqnarray}

Our analysis shows that $r^{(2)}_{ph} $ is always negative and
$r^{(3)}_{ph} $ given by Eq. (\ref{Eqroot3}) is the largest
positive root. From the definition of the shadow radius
\cite{Perlick2015}, the size of BH shadow can be obtained as
\begin{equation}
r_{sh}=\frac{L_{p}}{E}=\frac{r_{ph}}{\sqrt{f(r_{ph})}}.  \label{Eqrsh}
\end{equation}

The shape of the shadow seen by an observer at spatial infinity
can be obtained from the geodesics of the photons in the celestial
coordinates $x$ and $y$  as follows \cite{Vazquez1g}
\begin{eqnarray}
\label{celestial:1a}
x&=&\lim_{r_0\rightarrow\infty}\left(-r_0^2\sin\theta_0\frac{d\varphi}{dr}\Big|_{(r_0,\theta_0)}\right),\\ \nonumber
y&=&\lim_{r_0\rightarrow\infty}\left(r_0^2\frac{d\theta}{dr}\Big|_{(r_0,\theta_0)}\right),
\end{eqnarray}
where $ r_{0} $ is the distance between the observer and the BH,
and $ \theta_{0} $ is the inclination angle. For our solution, we
obtain that
\begin{equation}
x=-\frac{r_{sh}}{\sqrt{1+\frac{\Lambda}{3}r_{sh}^{2}}}, ~~~~y=0,
\end{equation}

It should be noted though the celestial coordinates are well
defined for $ r_{0} \longrightarrow \infty$ in our work, there are
some cases where the celestial coordinates are not well defined in
such a limit. For such cases, one can consider an observer at
"$r_{0} =$ constant" to determine the shape of a shadow (see Refs.
\cite{Perlick2018,Grenzebach1a} for more details).

\begin{figure}[H]
\centering \subfloat[$ Q=\varepsilon=0.2$ and $ \Lambda=-0.01 $]{
        \includegraphics[width=0.31\textwidth]{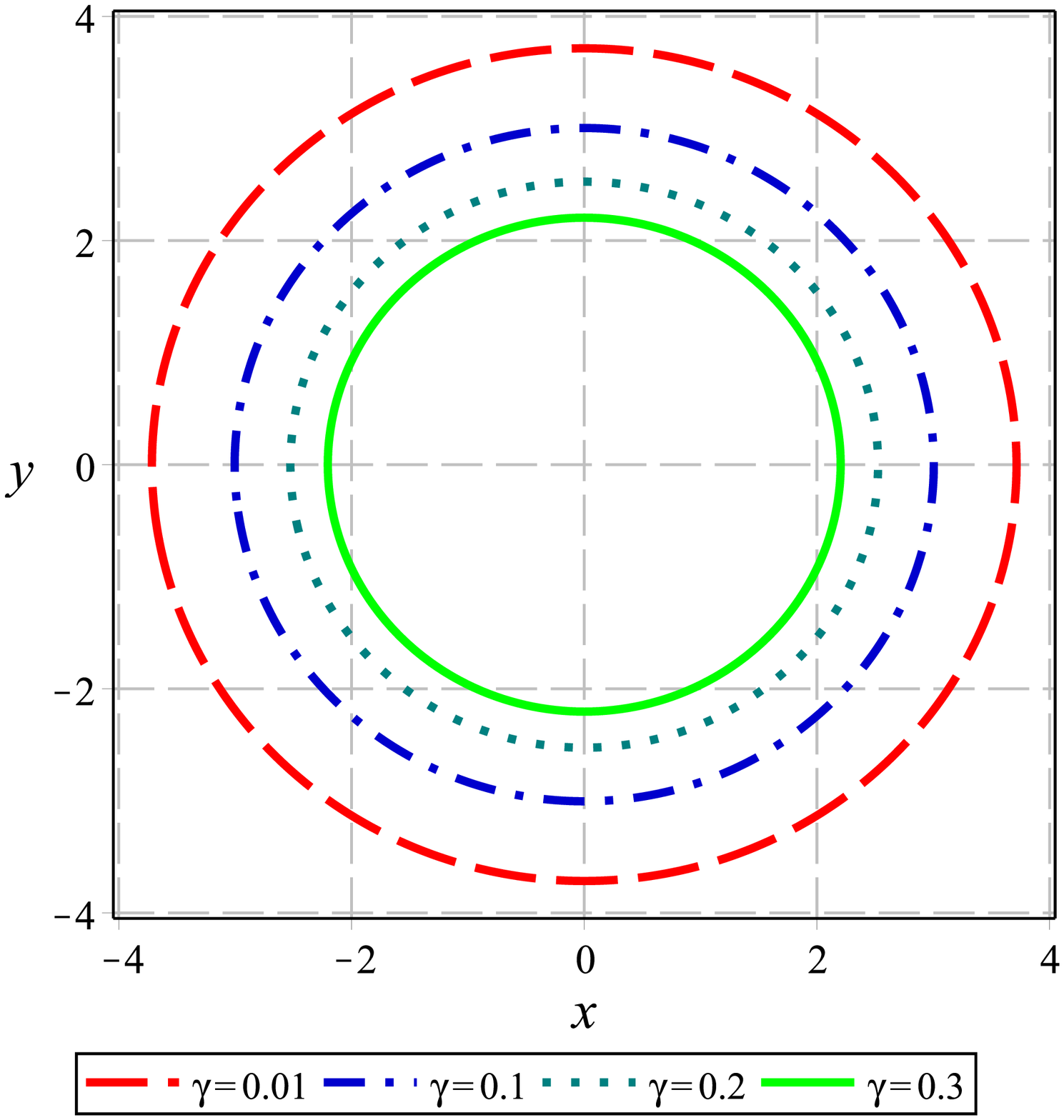}}
\subfloat[$  Q=\gamma =0.2$ and $ \Lambda=-0.01 $]{
        \includegraphics[width=0.315\textwidth]{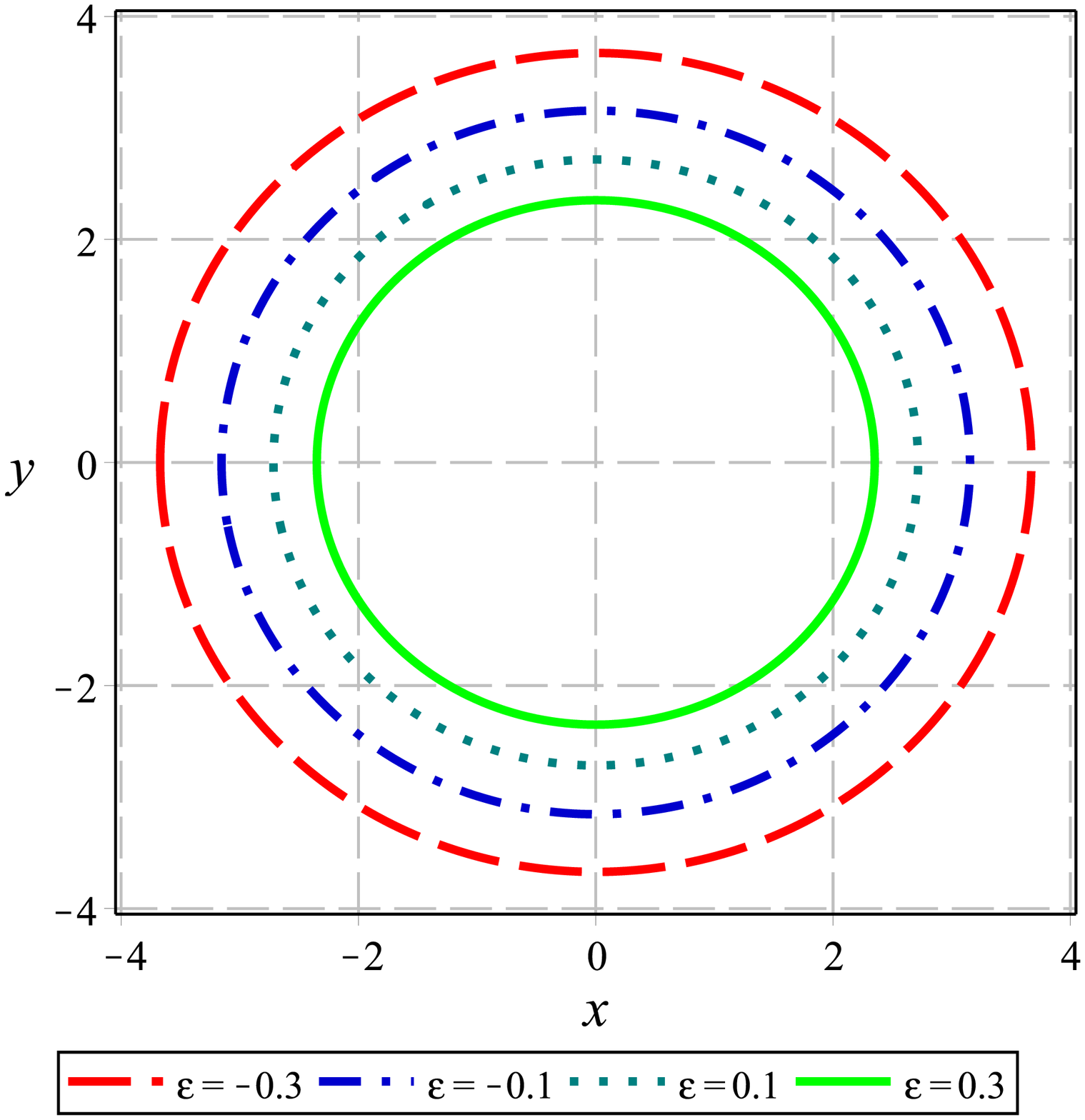}}\newline
\subfloat[ $ \gamma =\varepsilon=0.2$ and $ \Lambda=-0.01 $]{
        \includegraphics[width=0.31\textwidth]{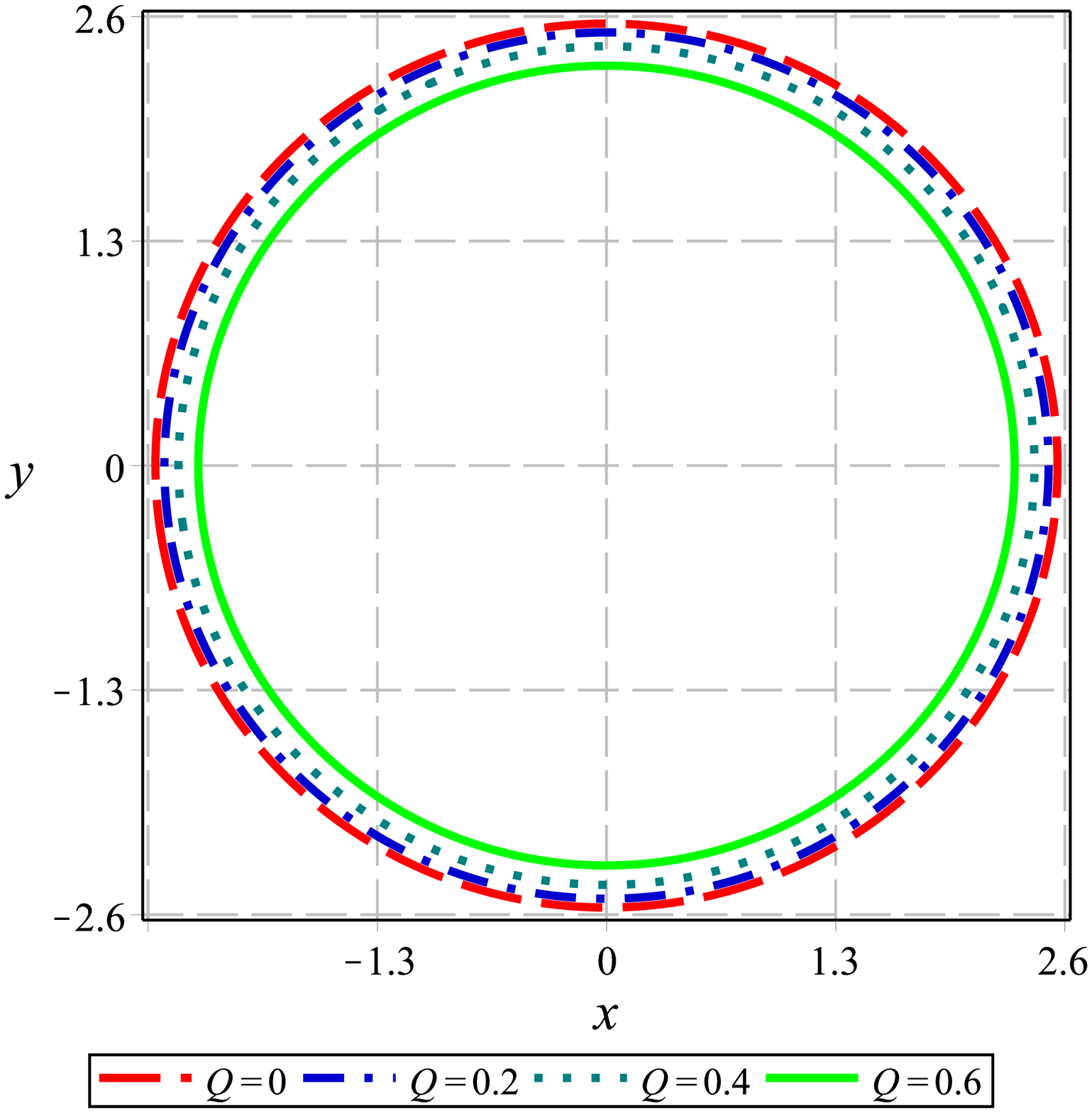}}
\subfloat[ $ Q= \gamma =\varepsilon=0.2$]{
        \includegraphics[width=0.31\textwidth]{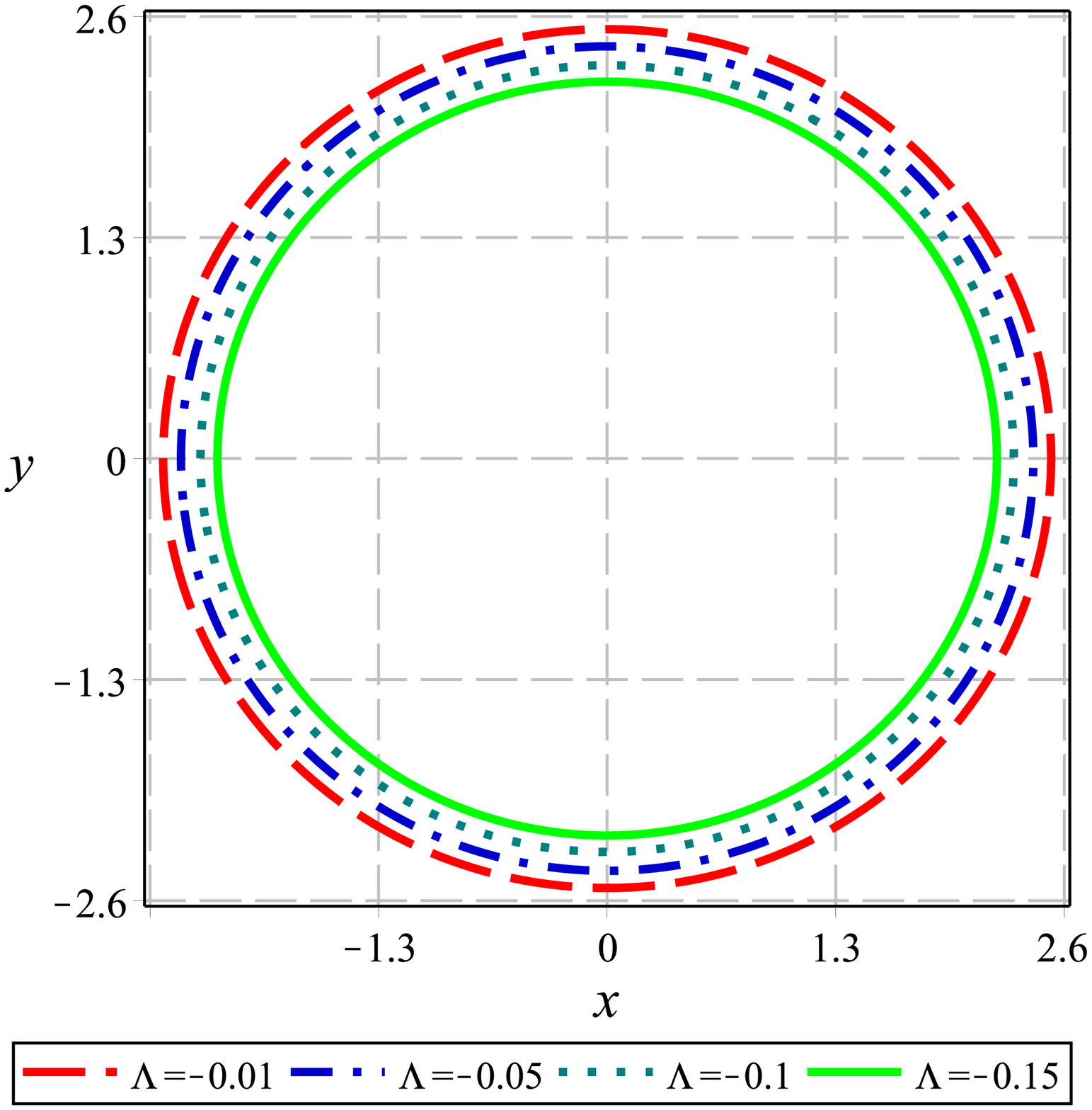}}\newline
\caption{The BH shadow in the celestial plane $(x-y)$ for $M=1$.}
\label{Fig4}
\end{figure}

\begin{table*}[htb!]
\caption{The event horizon ($r_{e}$), photon sphere radius ($r_{ph}$) and
shadow radius ($r_{sh}$) for the variation of $Q$, $\gamma$, $\varepsilon $ and $\Lambda$ for $M =1$.}
\label{table2}\centering
\begin{tabular}{||c|c|c|c|c||}
\hline
{\footnotesize $Q$ \hspace{0.3cm}} & \hspace{0.3cm}$0.2$ \hspace{0.3cm}
& \hspace{0.3cm} $0.4$\hspace{0.3cm} & \hspace{0.3cm} $0.6$\hspace{0.3cm} &
\hspace{0.3cm}$0.9$\hspace{0.3cm} \\ \hline
$r_{e}$ ($\gamma =\varepsilon=0.2$, $\Lambda =-0.1 $) & $1.2989$
& $1.2468$ & $1.1445$ & $0.69+0.2I$ \\ \hline
$r_{ph}$ ($\gamma =\varepsilon=0.2$, $\Lambda =-0.1 $) & $2.1005 $
& $2.0262$ & $1.8855$ & $1.3461$ \\ \hline
$r_{sh}$ ($\gamma =\varepsilon=0.2$, $\Lambda =-0.1 $) & $2.3138 $
& $2.2739$ & $2.1975$ & $1.9212$ \\ \hline
$r_{ph}>r_{e}$ & \checkmark & \checkmark & \checkmark & $\times$ \\ \hline
$r_{sh}>r_{ph}$ & \checkmark & \checkmark & \checkmark & \checkmark \\
\hline\hline
&  &  &  &  \\
{\footnotesize $\gamma$ \hspace{0.3cm}} & \hspace{0.3cm} $-0.1$ \hspace{0.3cm} &
\hspace{0.3cm} $0.2$\hspace{0.3cm} & \hspace{0.3cm} $0.3$\hspace{0.3cm} &
\hspace{0.3cm} $0.4$\hspace{0.3cm} \\ \hline
$r_{e}$ ($Q =\varepsilon=0.2$, $\Lambda =-0.1 $) & $1.7541$ & $1.2989$ & $1.2181$ & $1.1524$ \\ \hline
$r_{ph}$ ($Q =\varepsilon=0.2$, $\Lambda =-0.1 $) & $2.5037 $ & $2.1005$ & $1.9774$ & $1.8771$ \\ \hline
$r_{sh}$ ($Q =\varepsilon=0.2$, $\Lambda =-0.1 $) & $8.5704 $ & $2.3138$ & $2.0589$ & $1.8639$ \\ \hline
$r_{ph}>r_{e}$ & \checkmark & \checkmark & \checkmark & \checkmark \\ \hline
$r_{sh}>r_{ph}$ &  \checkmark & \checkmark & \checkmark & $\times$ \\
\hline\hline
&  &  &  &  \\
{\footnotesize $\varepsilon$ \hspace{0.3cm}} & \hspace{0.3cm}$-0.4$ \hspace{0.3cm} &
\hspace{0.3cm} $-0.1$\hspace{0.3cm} & \hspace{0.3cm} $0.6$\hspace{0.3cm} &
\hspace{0.3cm}$1.5$\hspace{0.3cm} \\ \hline
$r_{e}$ ($Q =\gamma=0.2$, $\Lambda =-0.1 $) & $1.8336 $ & $1.5353$ & $1.0608$ & $0.7047$ \\ \hline
$r_{ph}$ ($Q =\gamma=0.2$, $\Lambda =-0.1 $) & $3.2614 $ & $2.5667$ & $1.6706$ & $1.0804$ \\ \hline
$r_{sh}$ ($Q =\gamma=0.2$, $\Lambda =-0.1 $) & $3.0224 $ & $2.7680$ & $1.8181$ & $1.0603$ \\ \hline
$r_{ph}>r_{e}$ & \checkmark & \checkmark & \checkmark & \checkmark \\ \hline
$r_{sh}>r_{ph}$ & $\times$ & \checkmark & \checkmark & $\times$ \\
\hline\hline
&  &  & &  \\
{\footnotesize $\Lambda$ \hspace{0.3cm}} & \hspace{0.3cm}$0.01$ \hspace{%
0.3cm} & \hspace{0.3cm} $-0.04$\hspace{0.3cm} & \hspace{0.3cm} $-0.1$\hspace{%
0.3cm} & \hspace{0.3cm}$-0.21$\hspace{0.3cm} \\ \hline
$r_{e}$ ($Q =\gamma=\varepsilon=0.2$) & $65.36 $ & $1.3236$ & $1.2989$ & $1.2626$ \\ \hline
$r_{ph}$ ($Q =\gamma=\varepsilon=0.2$) & $2.1005 $ & $2.1005$ & $2.1005$ & $2.1005$ \\ \hline
$r_{sh}$ ($Q =\gamma=\varepsilon=0.2$) & $2.5810 $ & $2.4486$ & $2.3138$ & $2.0999$ \\ \hline
$r_{ph}>r_{e}$ & $\times$ & \checkmark & \checkmark & \checkmark \\ \hline
$r_{sh}>r_{ph}$ & \checkmark & \checkmark & \checkmark & $\times$ \\
\hline\hline
\end{tabular}%
\end{table*}
In order to observe an acceptable optical behavior, we need to
investigate the condition $r_{e}<r_{ph}<r_{sh}$, where $r_{e} $ is
radius related to the event horizon. It helps us find
admissible regions of $\gamma $ and $\varepsilon $ for having an
acceptable physical result. For clarity, we list several values of
the event horizon, the photon sphere radius, and shadow radius in
table \ref{table2}. As one can see, the increase of the electric
charge leads to an imaginary event horizon, indicating that an
acceptable optical result can be observed only for limited regions
of this parameter. Regarding the effects of the cosmological
constant and massive parameters $ \gamma $ and $ \varepsilon $, we
notice that for large values of these parameters, the shadow size
is smaller than the photon sphere radius which is not acceptable
physically. As a remarkable point regarding the parameter $ \gamma
$, we find that for negative values of this parameter, the shadow
radius becomes too large compared to the size of the photon sphere
radius ($r_{sh}\gg r_{ph} $). It reveals the fact that just for
positive values of $ \gamma $, one can obtain acceptable results.
From this table, one can also find that the electric charge and
massive parameters $ \gamma $ and $ \varepsilon $ have decreasing
effects on the event horizon, photon sphere radius, and shadow
size. Studying the effect of the cosmology constant, we find that
increasing this parameter leads to decreasing the event horizon
and shadow radii. Besides, it is clear that the cosmological
constant does not affect the photon sphere radius. It is worth
mentioning that according to our calculations, no acceptable
optical behavior is observed in dS spacetime. To have a precise
picture of the influence of these parameters on the shadow size,
we have plotted Fig. \ref{Fig4}. From this figure, one can find
that massive parameters have a significant effect on the size of
the black hole shadow compared to the electric charge and the
cosmological constant parameters.
\subsection{Energy emission}\label{Sec. IIB}

Having the BH shadow, one can investigate the emission of
particles around the BH. It was shown that the BH shadow
corresponds to its high energy absorption cross-section for a far
distant observer \cite{Wei2013}. In general, for a spherically
symmetric BH, the absorption cross-section oscillates around a
limiting constant value $\sigma _{lim}$ in the limit of very high
energies. Since the shadow measures the optical appearance of a
BH, it is approximately equal to the area of the photon sphere
($\sigma _{lim}\approx \pi r_{sh}^{2}$). The energy emission rate
is obtained as
\begin{equation}
\frac{d^{2}\mathcal{E}(\omega )}{dtd\omega }=\frac{2\pi ^{3}\omega
^{3}r_{sh}^{2}}{e^{\frac{\omega }{T}}-1},  \label{Eqemission}
\end{equation}%
in which $\omega $ is the emission frequency, and $T$ is Hawking temperature given by
\begin{equation}
T=\frac{-\Lambda r_{e}^{4}+2\gamma r_{e}^{3}+(1+\varepsilon )r_{e}^{2}-Q^{2}%
}{4\pi r_{e}^{3}},  \label{EqTH}
\end{equation}

\begin{figure}[H]
\centering \subfloat[$ Q=\varepsilon=0.2$ and $ \Lambda=-0.01 $]{
        \includegraphics[width=0.31\textwidth]{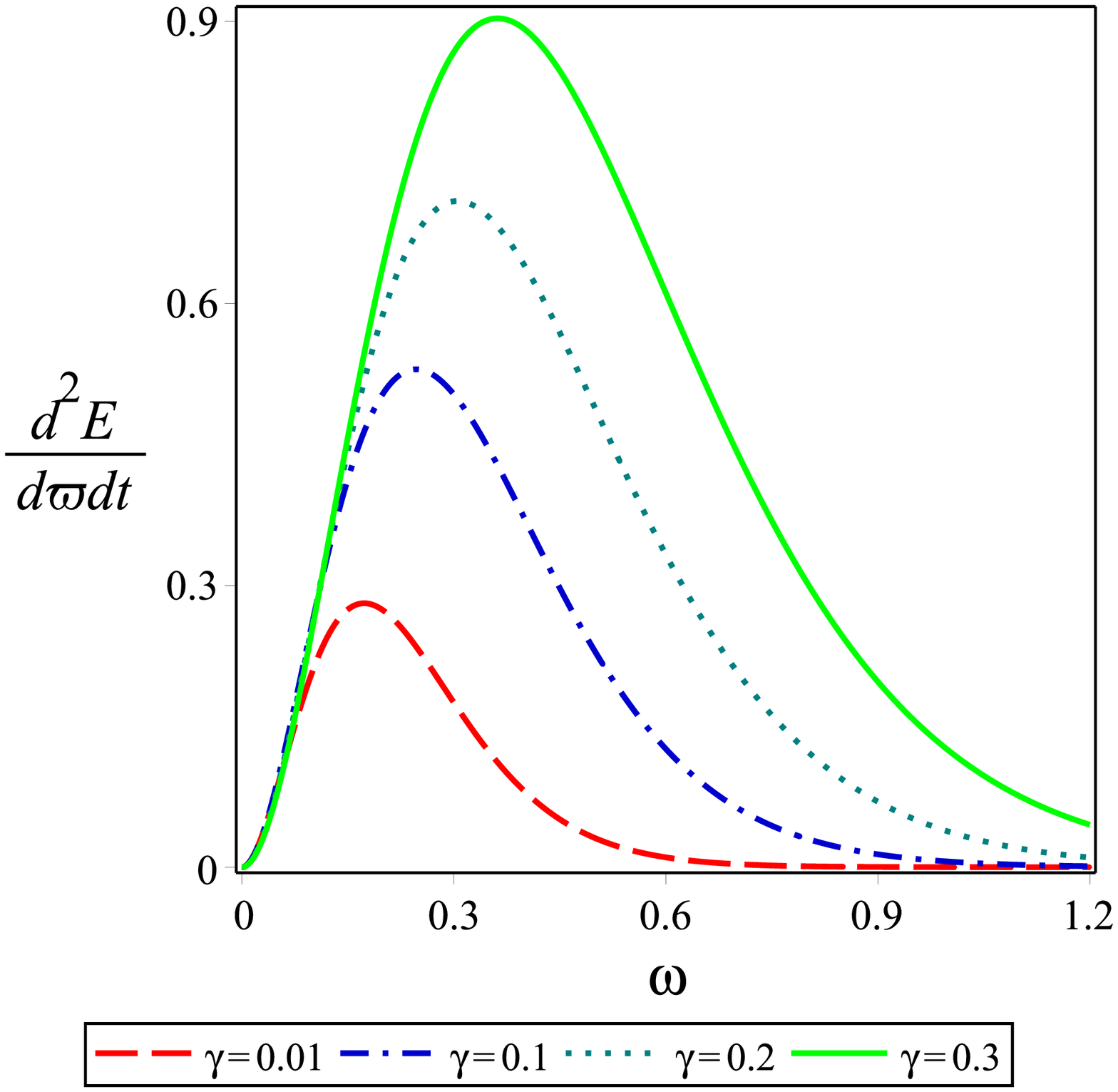}}
\subfloat[$  Q=\gamma =0.2$ and $ \Lambda=-0.01 $]{
        \includegraphics[width=0.31\textwidth]{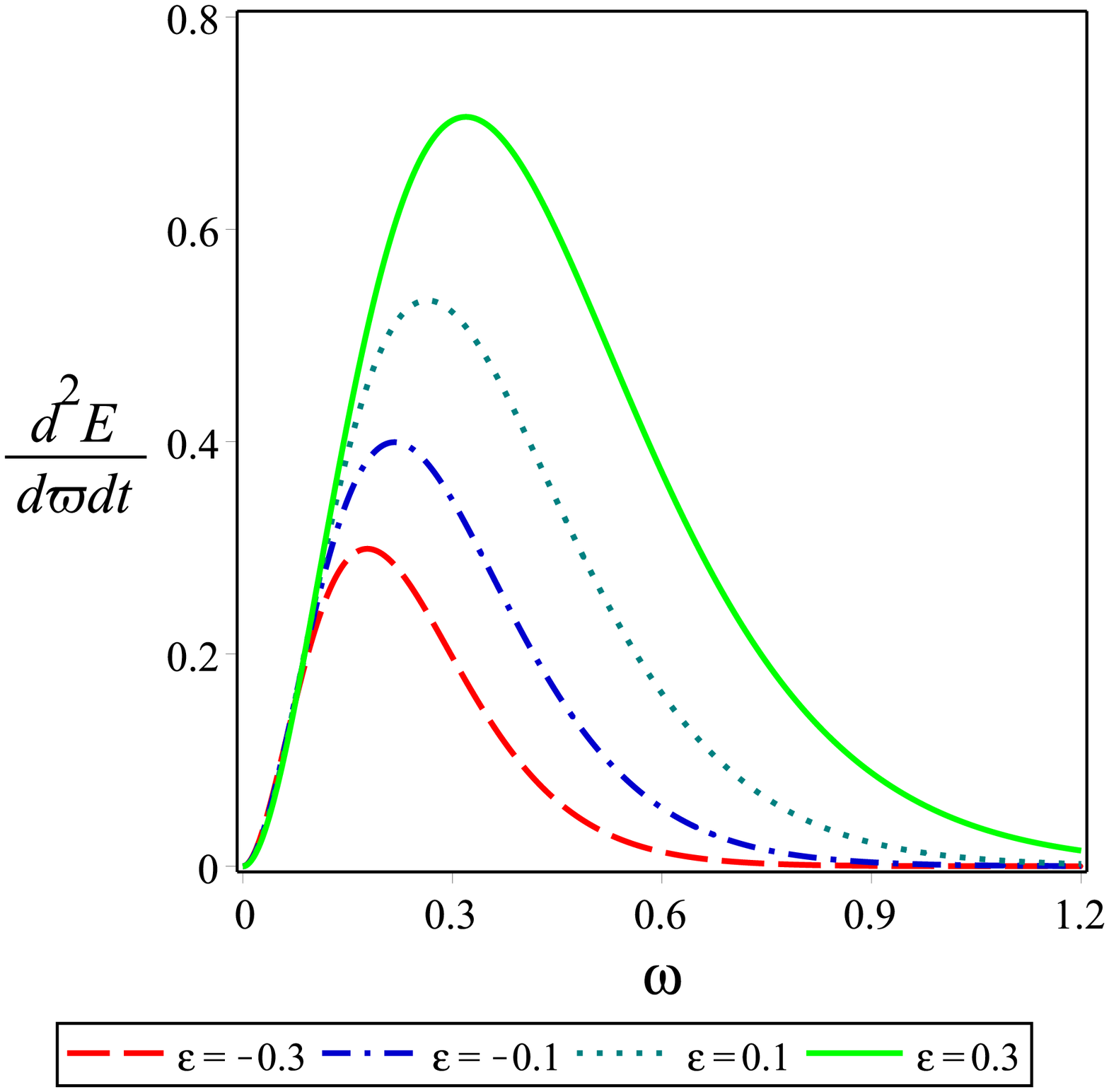}}\newline
\subfloat[ $ \gamma =\varepsilon=0.2$ and $ \Lambda=-0.01 $]{
        \includegraphics[width=0.31\textwidth]{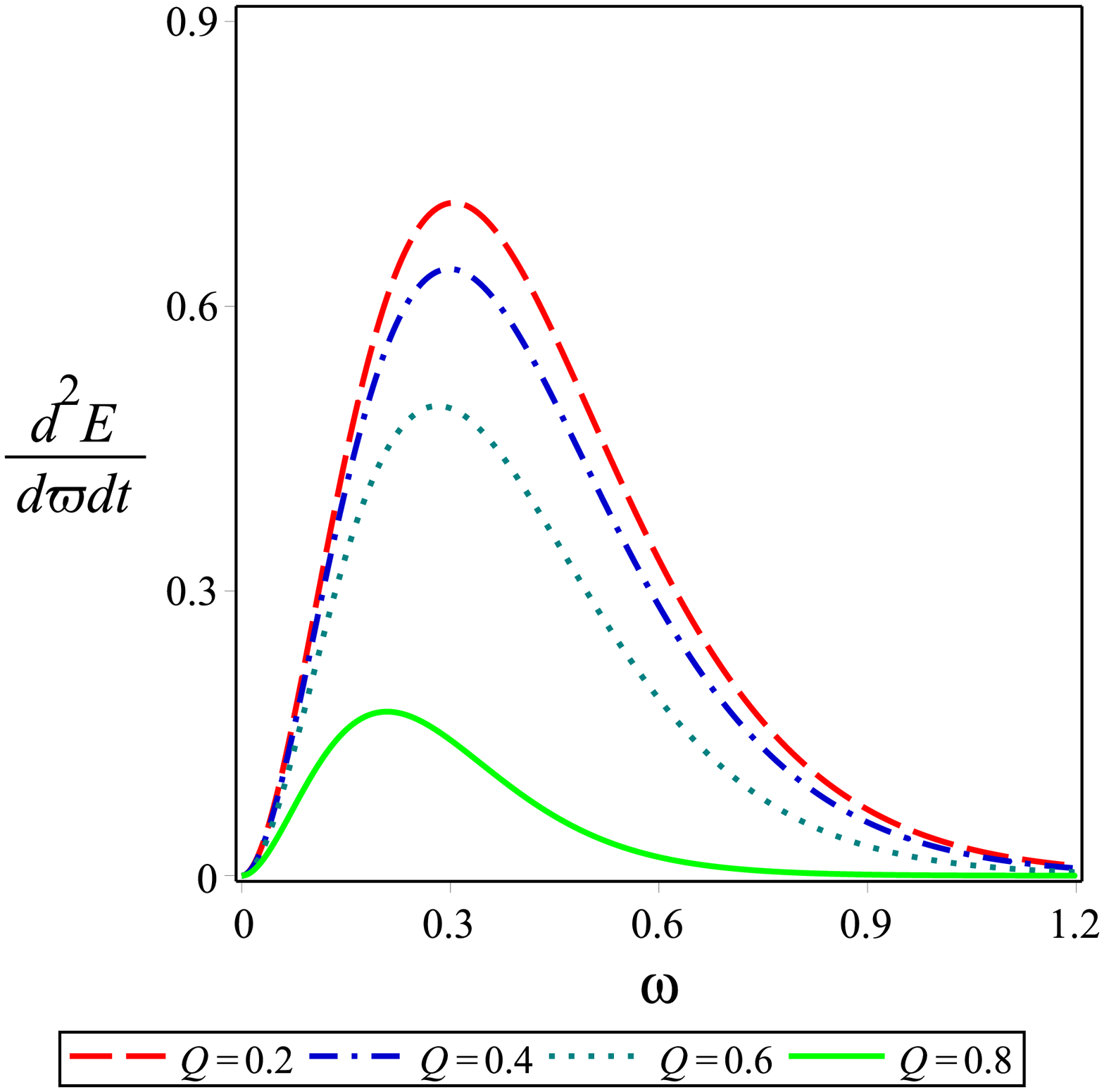}}
\subfloat[ $ Q= \gamma =\varepsilon=0.2$]{
        \includegraphics[width=0.31\textwidth]{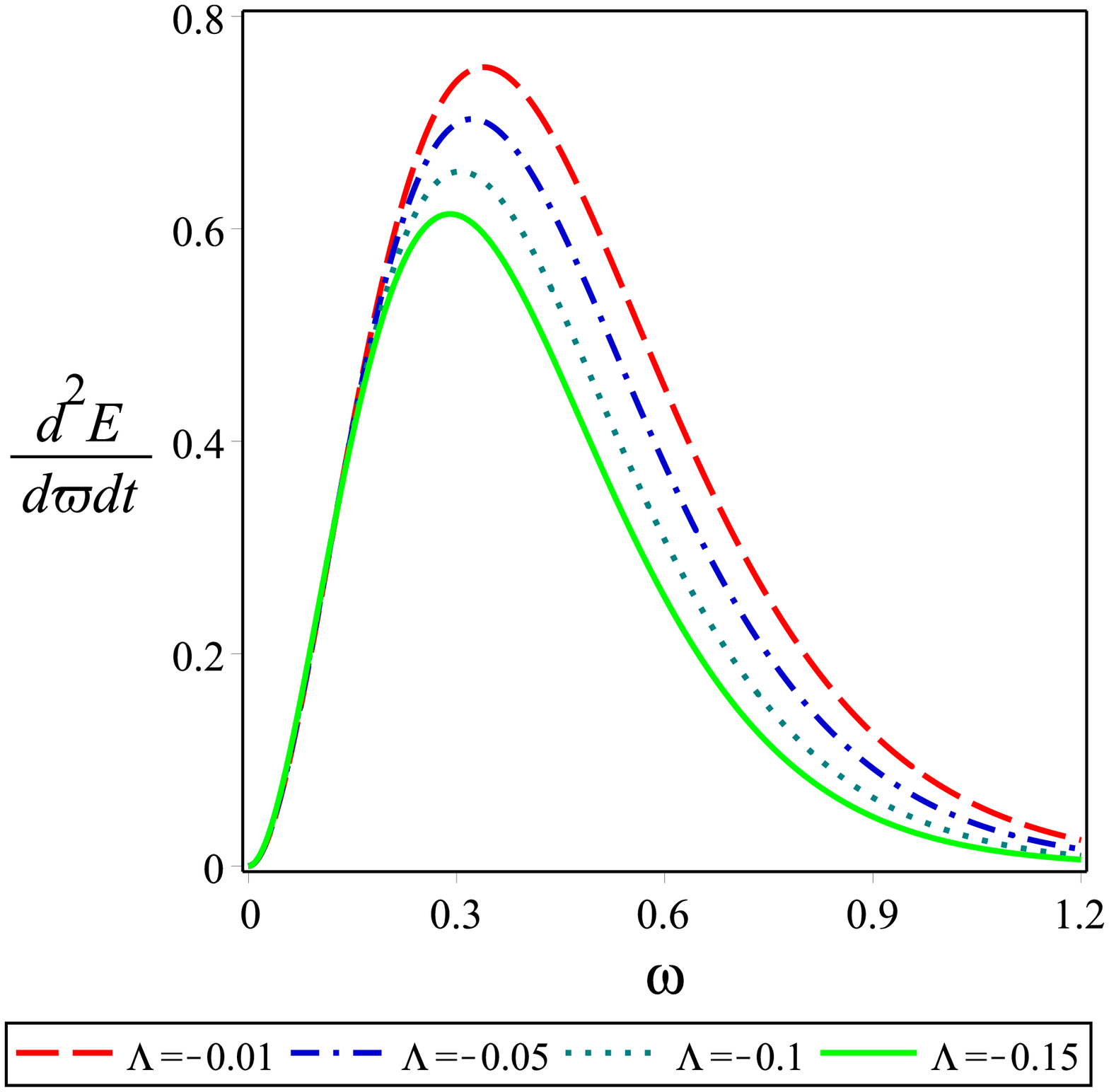}}\newline
\caption{The energy emission rate versus $\protect\omega $ for the
corresponding BH with $M=1$ and different values
of $\protect\gamma $, $\protect\varepsilon $, $Q$ and $\Lambda$.}
\label{Fig5}
\end{figure}

Figure \ref{Fig5}, displays the behavior of energy emission rate
as a function of $\omega $. As it is transparent, there exists a
peak of the energy emission rate, which decreases and shifts to
low frequencies as the parameters $\gamma $ and $\varepsilon $
decrease. Increasing these parameters leads to a fast emission of
particles. Regarding the effect of electric charge, as we see from
Fig. \ref{Fig5}(c), this parameter decreases the energy emission
so that when the BH is located in a weak electric field, the
evaporation process would be faster. Studying the impact of the
cosmological constant, we observe that $ \vert \Lambda \vert $ has
a decreasing effect on this optical quantity like the electric
charge (see Fig. \ref{Fig5}(d)). Therefore, one can find that the
black hole has a longer lifetime in a high curvature background or
with a strong electric field.
\section{Rotating charged BH in dRGT massive gravity}\label{Sec. III}

In this section, we would like to generalize the spherically
symmetric charged BH solution in dRGT massive gravity to the Kerr-like
rotational BH solution by employing the Newman-Janis algorithm
(NJA), which was first proposed by Newman and Janis in 1965
\cite{Newman1965}. This method has been widely used to construct
rotating BH solutions from their non-rotating counterparts
\cite{xu2017kerr,kim2020rotating}.
In particular, it has been more recently applied to non-rotating
solutions in modified gravity theories
\cite{jusufi2020rotating,Ghosh1ah}.
In GR, if a source exists, it is the same for both non-rotating BH
and its rotating counterpart, e.g., charge for
Reissner-Nordstr\"{o}m and Kerr-Newman BHs. However, in the
modified gravity,  rotating BH counterparts obtained using NJA, in
addition to original sources likely have additional sources. In
this work, we will adopt the NJA modified by Azreg-A\"inou
\cite{azreg2014generating,azreg2014static}, which can generate
rotating solutions without complication.

First, we consider the general static and spherically symmetric metric
 \begin{equation}
\text{d}s^{2}=-f(r)\text{d}t^{2}+g(r)^{-1}\text{d}r^{2}+h(r)\text{d}\Omega
^{2}.  \label{Metric2}
\end{equation}

Then, we transform the metric (\ref{Metric2}) from the
Boyer-Lindquist (BL) coordinates ($t, r, \theta, \varphi$) to the
Eddington-Finkelstein (EF) coordinates ($u, r, \theta, \varphi$).
This can be achieved by using the following coordinate
transformation
\begin{equation}
  \text{d}u=\text{d}t-\frac{\text{d}r}{\sqrt{f(r)g(r)}}.
\end{equation}

So, the line element (\ref{Metric2}) takes the form
  \begin{equation}\label{Metric3}
  \text{d}s^{2}=-f\left(r\right)\text{d}u^{2}-2\sqrt{\frac{f\left(r\right)}{g\left(r\right)}}\text{d}u\text{d}r+h\left(r\right)\left(\text{d}\theta^{2}+\sin^{2}{\theta}\text{d}\varphi^{2}\right).
  \end{equation}

This metric can be decomposed in terms of null tetrad formalism as
\begin{equation}\label{eq:null}
  g^{\mu\nu}=-l^{\mu}n^{\nu}-l^{\nu}n^{\mu}+m^{\mu}\bar{m}^{\nu}+m^{\nu}\bar{m}^{\mu},
  \end{equation}
where
  \begin{equation}
  \begin{gathered}
  l^{\mu}=\delta^{\mu}_{r},\qquad n^{\mu}=\sqrt{\frac{g\left(r\right)}{f\left(r\right)}}\delta^{\mu}_{0}-\frac{f\left(r\right)}{2}\delta^{\mu}_{r},\\
  m^{\mu}=\frac{1}{\sqrt{2h\left(r\right)}}\delta^{\mu}_{\theta}+\frac{i}{\sqrt{2h\left(r\right)}\sin{\theta}}\delta^{\mu}_{\varphi},\quad
  \bar{m}^{\mu}=\frac{1}{\sqrt{2h\left(r\right)}}\delta^{\mu}_{\theta}-\frac{i}{\sqrt{2h\left(r\right)}\sin{\theta}}\delta^{\mu}_{\varphi}.
  \end{gathered}
  \end{equation}

These vectors further satisfy the conditions for normalization, orthogonality and isotropy as
\begin{equation}
  \begin{gathered}
l^{\mu}l_{\mu}=n^{\mu}n_{\mu}=m^{\mu}m_{\mu}=\bar{m}^{\mu}\bar{m}_{\mu}=0,\\
l^{\mu}m_{\mu}=n^{\mu}m_{\mu}=l^{\mu}\bar{m}_{\mu}=n^{\mu}\bar{m}_{\mu}=0,\\
 l^{\mu}n_{\mu}=-m^{\mu}\bar{m}_{\mu}=1.
\end{gathered}
\end{equation}

Now, we consider complex coordinate transformations as follows
\begin{equation}
{x'}^{\mu} = x^{\mu} + ia (\delta_r^{\mu} - \delta_u^{\mu})
\cos\theta \rightarrow \\ \left\{\begin{array}{ll}
u' = u - ia\cos\theta, \\
r' = r + ia\cos\theta, \\
\theta' = \theta, \\
\varphi' = \varphi. \end{array}\right.
\end{equation}

We assume that the metric functions take a new form under these
transformations as: $f\left(r\right)\rightarrow F\left(r, a,
\theta\right)$, $g\left(r\right)\rightarrow G\left(r, a,
\theta\right)$, and
$h\left(r\right)\rightarrow\Sigma=r^{2}+a^{2}\cos^{2}{\theta}$
\cite{azreg2014generating,azreg2014static}. So, null tetrad basis
may take the new form
\begin{equation}
  \begin{split}
  l'^{\mu}&=\delta^{\mu}_{r},\quad n'^{\mu}=\sqrt{\frac{G}{F}}\delta^{\mu}_{0}-\frac{1}{2}F\delta^{\mu}_{r},\\
  m'^{\mu}&=\frac{1}{\sqrt{2\Sigma}}\left(\delta^{\mu}_{\theta}+ia\sin{\theta}\left(\delta^{\mu}_{0}-\delta^{\mu}_{r}\right)+\frac{i}{\sin{\theta}}\delta^{\mu}_{\varphi}\right),\\
  \bar{m}'^{\mu}&=\frac{1}{\sqrt{2\Sigma}}\left(\delta^{\mu}_{\theta}-ia\sin{\theta}\left(\delta^{\mu}_{0}-\delta^{\mu}_{r}\right)-\frac{i}{\sin{\theta}}\delta^{\mu}_{\varphi}\right).
  \end{split}
  \end{equation}

Using  Eq. (\ref{eq:null}), the contravariant components of the metric $g^{\mu\nu}$ is obtained as
\begin{equation}
  \begin{split}
  g^{uu}&=\frac{a^{2}\sin^{2}{\theta}}{\Sigma},\quad g^{rr}=G+\frac{a^{2}\sin^{2}{\theta}}{\Sigma},\\
  g^{\theta\theta}&=\frac{1}{\Sigma},\quad g^{\varphi\varphi}=\frac{1}{\Sigma\sin^{2}{\theta}},\\
  g^{ur}&=g^{ru}=-\sqrt{\frac{G}{F}}-\frac{a^{2}\sin^{2}{\theta}}{\Sigma},\\
  g^{u\varphi}&=g^{\varphi u}=\frac{a}{\Sigma},\quad g^{r\varphi}=g^{\varphi r}=-\frac{a}{\Sigma}.
  \end{split}
  \end{equation}

The new metric is found as follows
\begin{equation}\label{eq:rotating EF}
  \begin{split}
  \mathrm{d}s^{2}=&-F\mathrm{d}u^{2}-2\sqrt{\frac{F}{G}}\mathrm{d}u\mathrm{d}r+2a\left(F-\sqrt{\frac{F}{G}}\right)\sin^{2}{\theta}\mathrm{d}u\mathrm{d}\varphi+\Sigma\mathrm{d}\theta^{2}\\
  &+2a\sin^{2}{\theta}\sqrt{\frac{F}{G}}\mathrm{d}r\mathrm{d}\varphi+\sin^{2}{\theta}\left[\Sigma+a^{2}\left(2\sqrt{\frac{F}{G}}-F\right)\sin^{2}{\theta}\right]\mathrm{d}\varphi^{2}.
  \end{split}
  \end{equation}

Now, we revert Eq (\ref{eq:rotating EF})  back to BL coordinates by using the following transformation
\begin{equation}
  \mathrm{d}u=\mathrm{d}t+\lambda\left(r\right)\mathrm{d}r,\quad \mathrm{d}\varphi=\mathrm{d}\phi+\chi\left(r\right)\mathrm{d}r,
  \end{equation}
where
\begin{equation}
  \lambda\left(r\right)=-\frac{k(r)+a^{2}}{g\left(r\right)h\left(r\right)+a^{2}},\quad \chi\left(r\right)=-\frac{a}{g\left(r\right)h\left(r\right)+a^{2}}, \quad k\left(r\right)=\sqrt{\frac{g\left(r\right)}{f\left(r\right)}}h\left(r\right),
  \end{equation}
  with
 \begin{equation}
 F\left(r,\theta\right)=\frac{\left(g(r)h(r)+a^{2}\cos^{2}{\theta}\right)\Sigma}{\left(k(r)^{2}+a^{2}\cos^{2}{\theta}\right)^{2}}, \quad G\left(r,\theta\right)=\frac{g(r)h(r)+a^{2}\cos^{2}{\theta}}{\Sigma}.
 \end{equation}

  Hence the rotating counterpart corresponding to the metric (\ref{Metric2}) turns out to be

\begin{equation}\label{eq:rotating Kerr}
  \begin{split}
  \mathrm{d}s^{2}=&-\frac{\left(g(r)h(r)+a^{2}\cos^{2}{\theta}\right)\Sigma}{\left(k(r)+a^{2}\cos^{2}{\theta}\right)^{2}}\mathrm{d}t^{2}+\frac{\Sigma}{g(r)h(r)+a^{2}}\mathrm{d}r^{2}-2a\sin^{2}\theta\left[\frac{k(r)-g(r)h(r)}{\left(k(r)+a^{2}\cos^{2}{\theta}\right)^{2}}\right]\Sigma\mathrm{d}\phi\mathrm{d}t\\
  &+\Sigma\mathrm{d}\theta^{2}+\Sigma\sin^{2}\theta\left[1+a^{2}\sin^{2}\theta\frac{2k(r)-g(r)h(r)+a^{2}\cos^{2}\theta}{\left(k(r)+a^{2}\cos^{2}{\theta}\right)^{2}}\right]\mathrm{d}\phi^{2}.
  \end{split}
  \end{equation}

Comparing the line elements (\ref{Metric}) with (\ref{Metric2}),
one  finds $ f(r)=g(r) $ and $ h(r)=k(r)=r^{2} $. Finally, we
obtain the metric of charged rotating BHs in dRGT massive gravity
in the form
  \begin{equation}\label{eq:rotating metric}
  \begin{split}
  \mathrm{d}s^{2}=&-\left(1-\frac{2\rho r}{\Sigma}\right)\mathrm{d}t^{2}+\frac{\Sigma}{\Delta_{r}}\mathrm{d}r^{2}+\Sigma\mathrm{d}\theta^{2}-\frac{4a\rho r\sin^{2}{\theta}}{\Sigma}\mathrm{d}t\mathrm{d}\phi\\&+\sin^{2}{\theta}\left(r^{2}+a^{2}+\frac{2a^{2}\rho r\sin^{2}{\theta}}{\Sigma}\right)\mathrm{d}\phi^{2},
  \end{split}
  \end{equation}
  with
  \begin{equation}\label{eq:line element terms}
  \begin{gathered}
  \Delta_{r}=r^{2}+a^{2}-2Mr+Q^{2}-\frac{\Lambda}{3}r^{4}+\gamma r^{3}+\varepsilon r^{2},\\
  2\rho=2M-\frac{Q^{2}}{r}+\frac{\Lambda}{3}r^{3}-\gamma r^{2}-\varepsilon r,\\
  \Sigma=r^{2}+a^{2}\cos^{2}{\theta}.
  \end{gathered}
  \end{equation}
\subsection{Null geodesics}
\label{Sec. IIIA} In order to find the contour of a BH shadow
related to the rotating spacetime metric (\ref{eq:rotating
metric}), we need to separate the null geodesic equations using
the Hamilton-Jacobi equation given by

\begin{equation}
\frac{\partial S}{\partial \sigma}=- \frac{1}{2}g^{\mu\nu}\frac{\partial S}{\partial x^{\mu}}\frac{\partial S}{%
\partial x^{\nu}},
\label{HJ1}
\end{equation}
where $S $ and $\sigma $ are the Jacobi action and affine
parameter along the geodesics, respectively. Using known constants
of the motion, one can separate the Jacobi function as follows

\begin{equation}
S=-\frac{1}{2}\mu^2\sigma+Et-L\phi+S_r(r)+S_{\theta}(\theta),
\label{HJ2}
\end{equation}
where $ \mu $ is the mass of the test particle, $ E=p_{t} $ and $
L= p_{\phi}$ are, respectively, the conserved energy and angular
momentum. The functions $S_r(r)$ and $S_\theta(\theta)$
respectively depends on coordinates $r$ and $\theta$. Combining
Eq. (\ref{HJ1}) and Eq. (\ref{HJ2}) and applying the variable
separable method, we get the null geodesic equations for a moving
photon ($\mu=0$) around the rotating BH in dRGT massive gravity as
\begin{align}
\label{HJ3}
&\Sigma\frac{dt}{d\sigma}=\frac{r^2+a^2}{\Delta_{r}}[E(r^2+a^2)-aL]-a(aE\sin^2\theta-L),\\
&\Sigma\frac{dr}{d\sigma}=\sqrt{\mathcal{R}(r)},\\
&\Sigma\frac{d\theta}{d\sigma}=\sqrt{\Theta(\theta)},\\
\label{HJ4}
&\Sigma\frac{d\phi}{d\sigma}=\frac{a}{\Delta_{r}}[E(r^2+a^2)-aL]-\left(aE-\frac{L}{\sin^2\theta}\right),
\end{align}
where $\mathcal{R}(r)$ and $\Theta(\theta)$ read as
\begin{align}
\label{HJ5}
&\mathcal{R}(r)=[E(r^2+a^2)-aL]^2-\Delta_{r}[(aE-L)^2+\mathcal{K}],\\
&\Theta(\theta)=\mathcal{K}-\left(  \dfrac{L^2}{\sin^2\theta}-a^2E^2  \right) \cos^2\theta,
\end{align}
with $\mathcal{K}$ the Carter constant. For $\mathcal{K}=0$,
$\theta$-motion is suppressed, and all photon orbits are
restricted only to a plane ($ \theta=\pi/2 $), yielding unstable
circular orbits at the equatorial plane. In order to obtain the
size and shape of the BH shadow, we need to express the radial
geodesic equation in terms of the effective potential
$V_{\text{eff}}$ as
\begin{equation*}
\Sigma^2\left(\frac{dr}{d\sigma}\right)^2+V_{\text{eff}}=0.
\end{equation*}

Introducing two impact parameters $\xi$ and $\eta$ \cite{Chandrasekhar1g} as
\begin{equation}
\xi=L/E,  \;\; \;\; \;\;   \eta=\mathcal{K}/E^2.
\end{equation}

The effective potential $ V_{\text{eff}} $ reads
\begin{equation}\label{veff}
V_{\text{eff}}=\Delta_{r}((a-\xi)^2+\eta)-(r^2+a^2-a\;\xi)^2,
\end{equation}
where we have replaced $V_{\text{eff}}/E^2$ by $V_{\text{eff}}$. The boundary of the shadow is mainly determined by the
circular photon orbit, which satisfies the following conditions
\begin{equation}\label{cond}
V_{\text{eff}}(r_{ph})=0,\quad~~~\frac{dV_{\text{eff}}(r_{ph})}{dr}=0,
\end{equation}
whereas instability of orbits obeys condition
\begin{equation}
\frac{d^{2}V_{\text{eff}}(r_{ph})}{dr^{2}}<0,
\end{equation}

Solving  Eq. (\ref{cond}), the critical values of impact
parameters for photons unstable orbits read
\begin{eqnarray}
\xi_{c}&=& \frac{(r_{ph}^{2}+a^{2})\Delta'(r_{ph})-4r_{ph}\Delta (r_{ph})}{a\Delta'(r_{ph})},\\
\eta_{c}&=& \frac{r_{ph}^{2}\left[ 16a^{2}\Delta (r_{ph})+8r_{ph}\Delta (r_{ph})\Delta'(r_{ph})-16\Delta (r_{ph})^{2}-r_{ph}^{2}\Delta'(r_{ph})^{2}\right] }{a^{2}\Delta'(r_{ph})^{2}}.
\end{eqnarray}

\begin{table*}[htb!]
\caption{The event horizon ($r_{e}$), photon sphere radius ($r_{ph}$) and
shadow radius ($r_{sh}$) for the variation of $Q$, $a$, $\gamma$, $\varepsilon $ and $\Lambda$ for $M =1$ and $ \theta_{0} =0$.}
\label{table31}\centering
\begin{tabular}{||c|c|c|c|c||}
\hline
{\footnotesize $Q$ \hspace{0.3cm}} & \hspace{0.3cm}$0.1$ \hspace{0.3cm}
& \hspace{0.3cm} $0.3$\hspace{0.3cm} & \hspace{0.3cm} $0.5$\hspace{0.3cm} &
\hspace{0.3cm}$0.7$\hspace{0.3cm} \\ \hline
$r_{e}$ ($a=0.5 $, $\varepsilon =0.5$, $\gamma=0.2$, $\Lambda =-0.01 $) & $1.0228$
& $0.9721$ & $0.8414$ & $0.59+0.25I$ \\ \hline
$r_{ph}$ ($a=0.5 $, $\varepsilon =0.5$, $\gamma=0.2$, $\Lambda =-0.01 $) & $1.7395 $
& $1.6833$ & $1.5535$ & $1.2523$ \\ \hline
$r_{sh}$ ($a=0.5 $, $\varepsilon =0.5$, $\gamma=0.2$, $\Lambda =-0.01 $) & $2.1132 $
& $2.0768$ & $1.9949$ & $1.8285$ \\ \hline
$r_{ph}>r_{e}$ & \checkmark & \checkmark & \checkmark & $\times$ \\ \hline
$r_{sh}>r_{ph}$ & \checkmark & \checkmark & \checkmark & \checkmark \\
\hline\hline
&  &  &  &  \\
{\footnotesize $a$ \hspace{0.3cm}} & \hspace{0.3cm} $0.1$ \hspace{0.3cm} &
\hspace{0.3cm} $0.3$\hspace{0.3cm} & \hspace{0.3cm} $0.5$\hspace{0.3cm} &
\hspace{0.3cm} $0.7$\hspace{0.3cm} \\ \hline
$r_{e}$ ($Q=0.5 $, $\varepsilon =0.5$, $\gamma=0.2$, $\Lambda =-0.01 $) & $1.0228$ & $0.9721$ & $0.8414$ & $0.59+0.25I$ \\ \hline
$r_{ph}$ ($Q=0.5 $, $\varepsilon =0.5$, $\gamma=0.2$, $\Lambda =-0.01 $) & $1.6170 $ & $1.5975$ & $1.5535$ & $1.4674$ \\ \hline
$r_{sh}$ ($Q=0.5 $, $\varepsilon =0.5$, $\gamma=0.2$, $\Lambda =-0.01 $) & $1.9474 $ & $1.9637$ & $1.9949$ & $2.0378$ \\ \hline
$r_{ph}>r_{e}$ & \checkmark & \checkmark & \checkmark &$\times$ \\ \hline
$r_{sh}>r_{ph}$ & \checkmark & \checkmark & \checkmark & \checkmark \\
\hline\hline
&  &  &  &  \\
{\footnotesize $\gamma$ \hspace{0.3cm}} & \hspace{0.3cm}$0.1$ \hspace{0.3cm} &
\hspace{0.3cm} $0.3$\hspace{0.3cm} & \hspace{0.3cm} $0.5$\hspace{0.3cm} &
\hspace{0.3cm}$0.7$\hspace{0.3cm} \\ \hline
$r_{e}$ ($a=Q=0.5 $, $\varepsilon =0.5$, $\Lambda =-0.01 $) & $0.9102 $ & $0.7842$ & $0.6910$ & $0.6139$ \\ \hline
$r_{ph}$ ($a=Q=0.5 $, $\varepsilon =0.5$, $\Lambda =-0.01 $) & $1.6184 $ & $1.4998$ & $1.4151$ & $1.3764$ \\ \hline
$r_{sh}$ ($a=Q=0.5 $, $\varepsilon =0.5$, $\Lambda =-0.01 $) & $2.2413 $ & $1.8116$ & $1.5529$ & $1.3509$ \\ \hline
$r_{ph}>r_{e}$ & \checkmark & \checkmark & \checkmark & \checkmark \\ \hline
$r_{sh}>r_{ph}$ & \checkmark  & \checkmark & \checkmark & $\times$ \\
\hline\hline
&  &  &  &  \\
{\footnotesize $\varepsilon$ \hspace{0.3cm}} & \hspace{0.3cm}$-1.2$ \hspace{%
0.3cm} & \hspace{0.3cm} $-0.5$\hspace{0.3cm} & \hspace{0.3cm} $0.5$\hspace{%
0.3cm} & \hspace{0.3cm}$1$\hspace{0.3cm} \\ \hline
$r_{e}$ ($a=Q=0.5 $, $\gamma =0.2$, $\Lambda =-0.01 $) & $3.4719 $ & $1.9359$ & $0.8414$ & $0.4+0.09I$ \\ \hline
$r_{ph}$ ($a=Q=0.5 $, $\gamma =0.2$, $\Lambda =-0.01 $) & $6.7261 $ & $3.3498$ & $1.5535$ & $1.1639$ \\ \hline
$r_{sh}$ ($a=Q=0.5 $, $\gamma =0.2$, $\Lambda =-0.01 $) & $6.3948 $ & $4.2316$ & $1.9949$ & $1.4630$ \\ \hline
$r_{ph}>r_{e}$ & \checkmark & \checkmark & \checkmark & $\times$ \\ \hline
$r_{sh}>r_{ph}$ & $\times$ & \checkmark & \checkmark & \checkmark \\
\hline\hline
&  &  &  &  \\
{\footnotesize $\Lambda$ \hspace{0.3cm}} & \hspace{0.3cm}$-0.01$ \hspace{%
0.3cm} & \hspace{0.3cm} $-0.1$\hspace{0.3cm} & \hspace{0.3cm} $-0.2$%
\hspace{0.3cm} & \hspace{0.3cm}$-0.4$\hspace{0.3cm} \\ \hline
$r_{e}$ ($a=Q=0.5 $, $\gamma =0.2$, $\varepsilon =0.5 $) & $0.8414 $ & $0.8263$ & $0.8108$ & $0.7832$ \\ \hline
$r_{ph}$ ($a=Q=0.5 $, $\gamma =0.2$, $\varepsilon =0.5 $) & $1.5535 $ & $1.5701$ & $1.5887$ & $1.6692$ \\ \hline
$r_{sh}$ ($a=Q=0.5 $, $\gamma =0.2$, $\varepsilon =0.5 $) & $1.9949 $ & $1.9036$ & $1.8152$ & $1.6267$ \\ \hline
$r_{ph}>r_{e}$ & \checkmark & \checkmark & \checkmark & \checkmark \\ \hline
$r_{sh}>r_{ph}$ & \checkmark & \checkmark & \checkmark & $\times$ \\
\hline\hline
\end{tabular}%
\end{table*}

 Using Eq. (\ref{celestial:1a}) and the null
geodesic equations (\ref{HJ3})-(\ref{HJ4}), we get the relations
between celestial coordinates and impact parameters as
\begin{equation}
x=\frac{a\sin\theta_0 -\xi\csc\theta_0}{\sqrt{1+\frac{\Lambda}{3}\left[(a-\xi)^{2}+\eta \right] }}\,,\qquad
y=\pm \sqrt{\frac{\eta+a^2\cos^2\theta_0-\xi^2\cot^2\theta_0}{1+\frac{\Lambda}{3}\left[(a-\xi)^{2}+\eta \right] }}\,.\label{xy}
\end{equation}
\begin{figure}[H]
\centering
\subfloat[$ Q=\gamma =\varepsilon=0.2$, $ a=0.8$ and $ \Lambda=-0.01 $]{
        \includegraphics[width=0.325\textwidth]{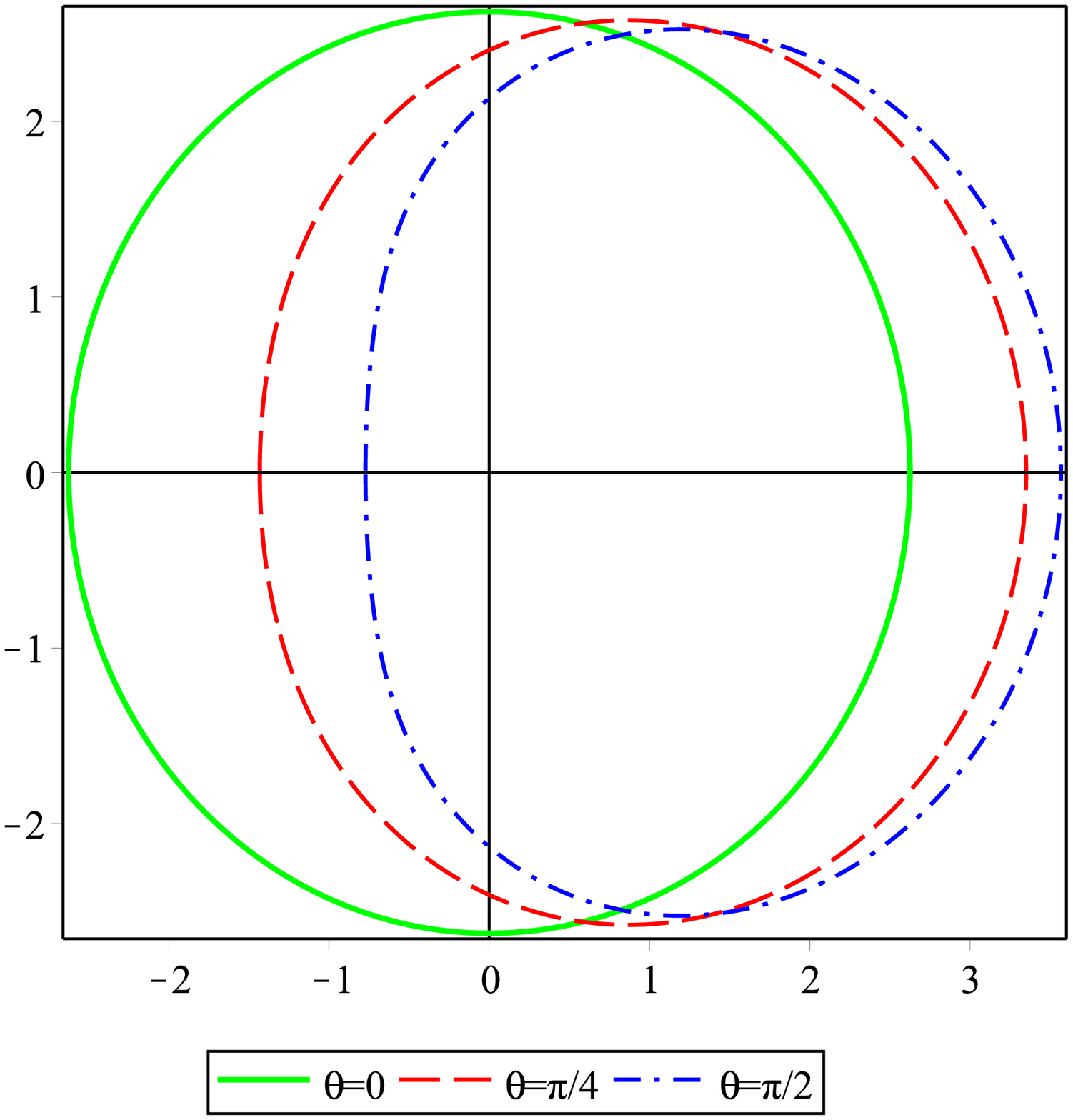}}
\subfloat[$ Q=\gamma =\varepsilon=0.2$, $ \Lambda=-0.01 $ and $ \theta_{0} =\pi/2$]{
        \includegraphics[width=0.32\textwidth]{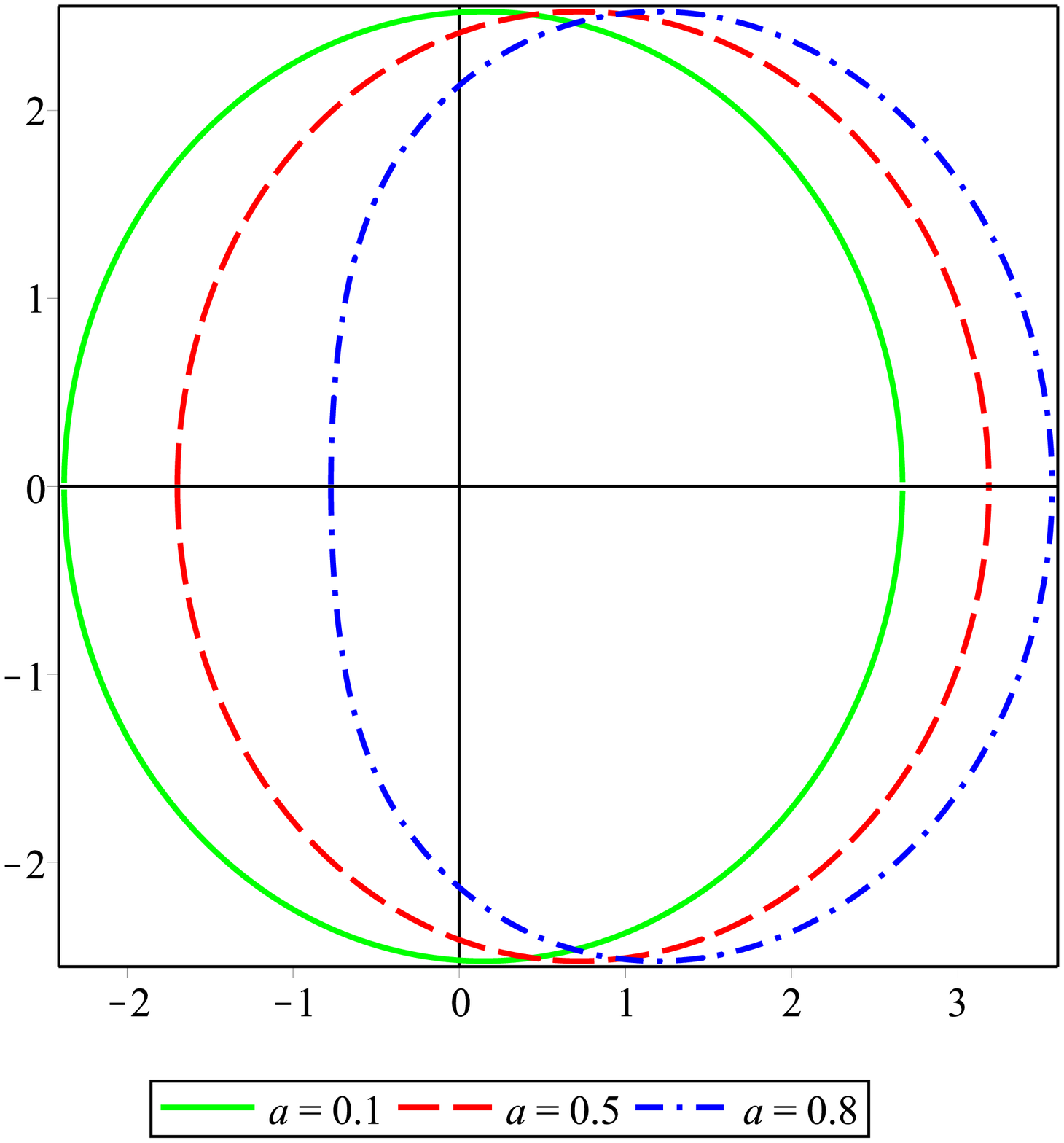}}
\subfloat[$ a=0.5 $, $ \gamma =\varepsilon=0.2$,  $ \Lambda=-0.01 $ and $ \theta_{0} =\pi/2$]{
        \includegraphics[width=0.306\textwidth]{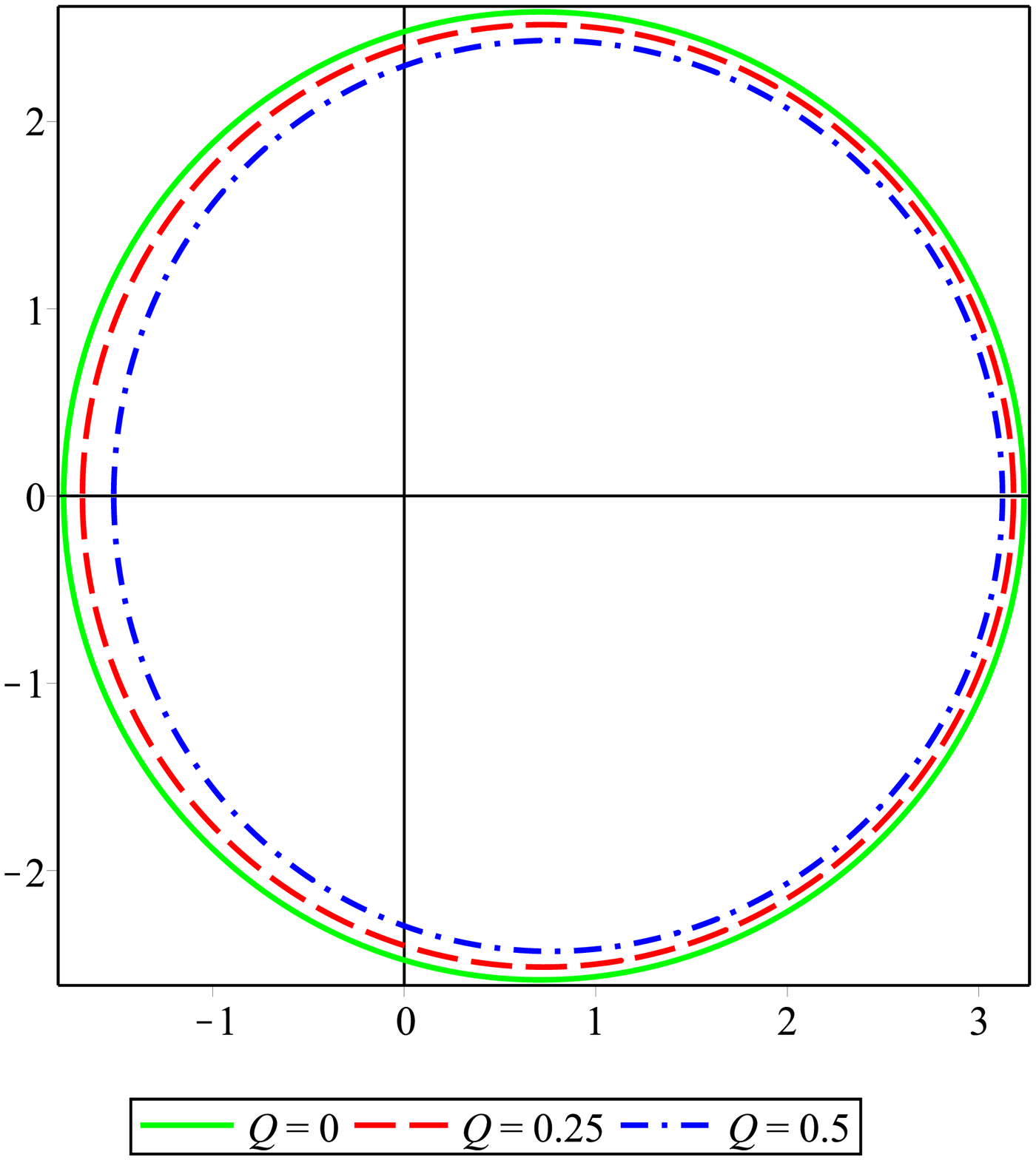}}\newline
\subfloat[$ Q=\varepsilon=0.2 $, $ a=0.5$, $ \Lambda=-0.01 $ and $ \theta_{0} =\pi/2$]{
        \includegraphics[width=0.33\textwidth]{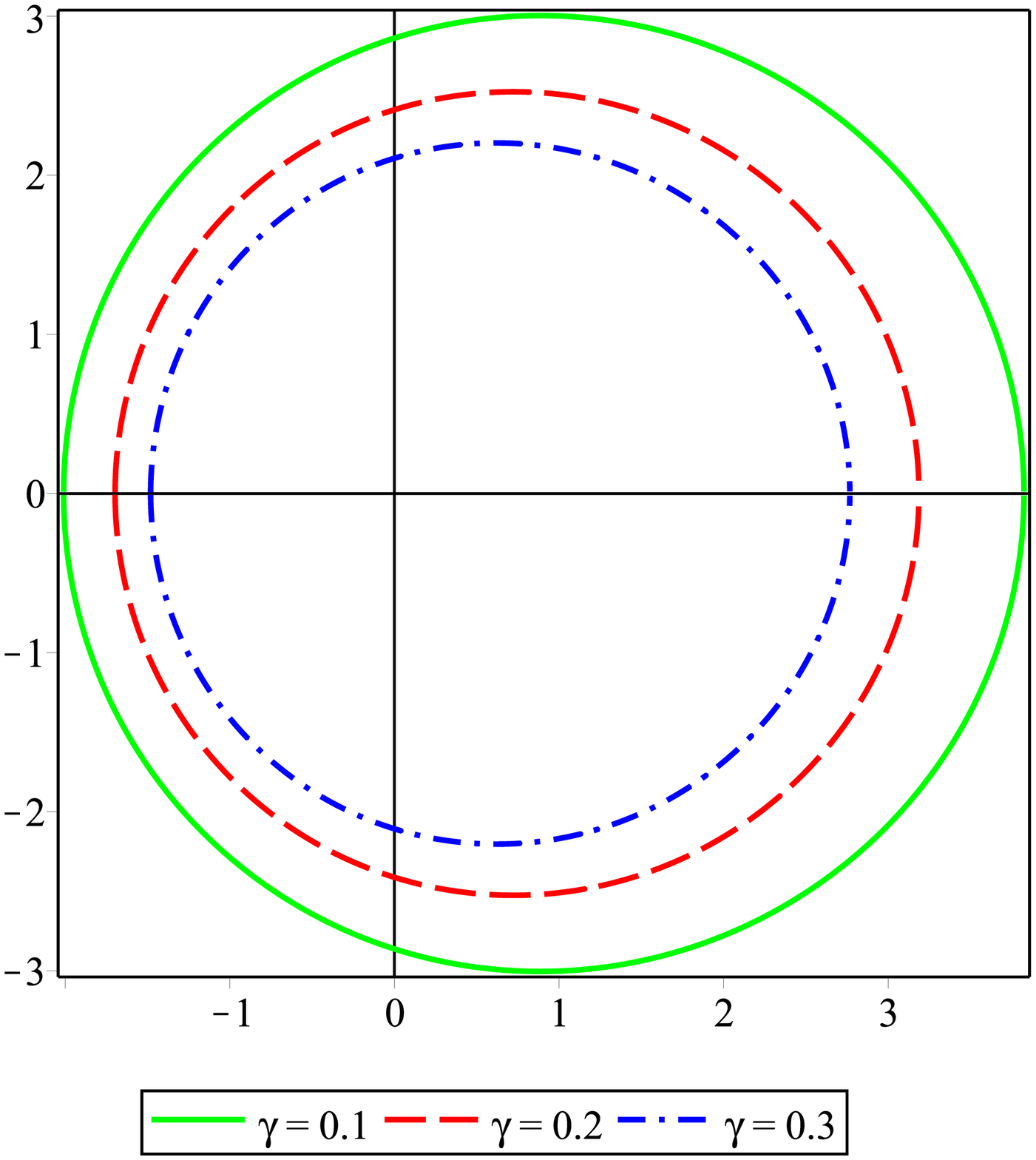}}
\subfloat[$ Q=\gamma=0.2 $, $ a=0.5$, $ \Lambda=-0.01 $ and $ \theta_{0} =\pi/2$]{
        \includegraphics[width=0.33\textwidth]{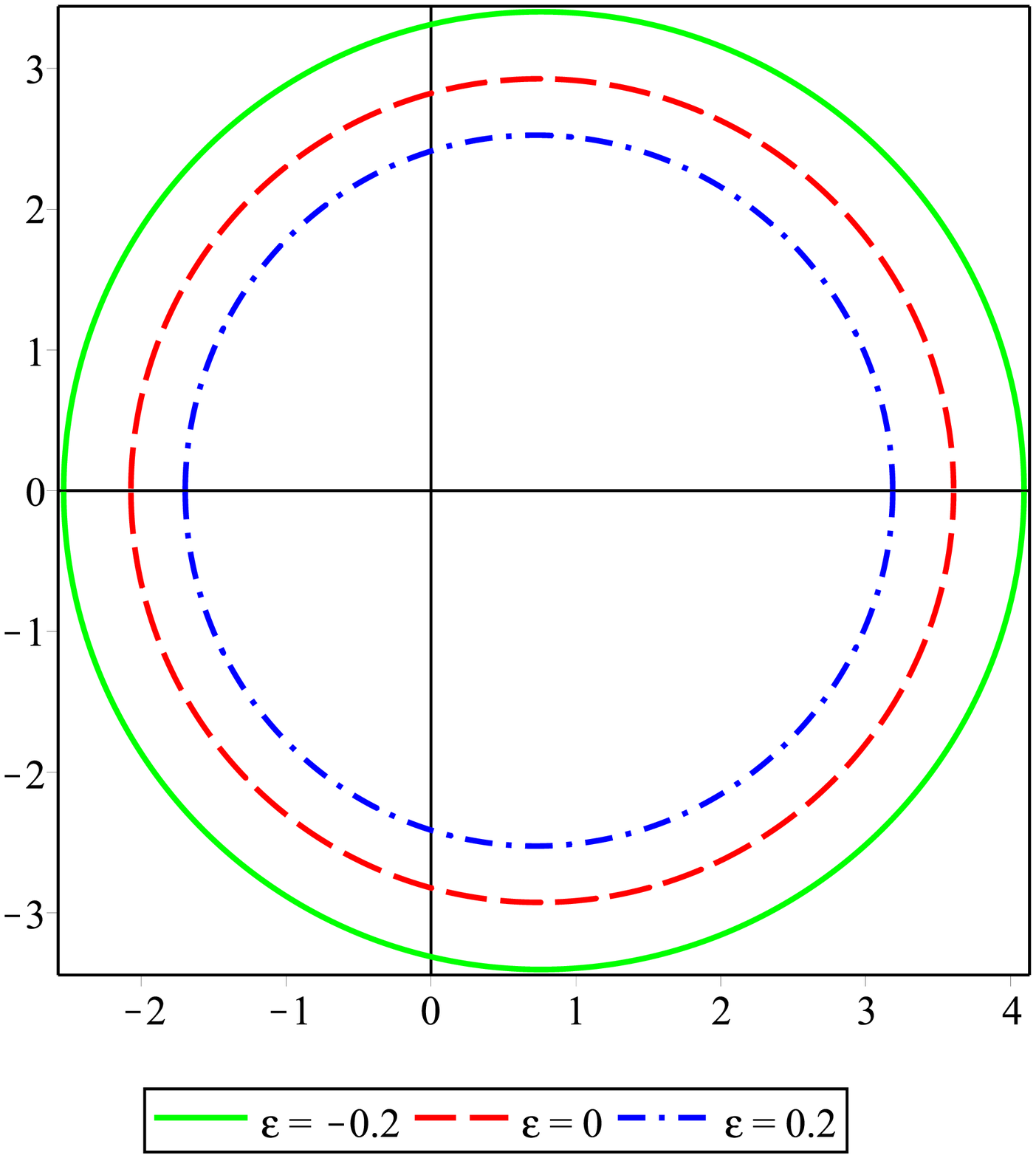}}
\subfloat[$ Q=\gamma=\varepsilon=0.2 $, $ a=0.5$ and $ \theta_{0} =\pi/2$]{
        \includegraphics[width=0.33\textwidth]{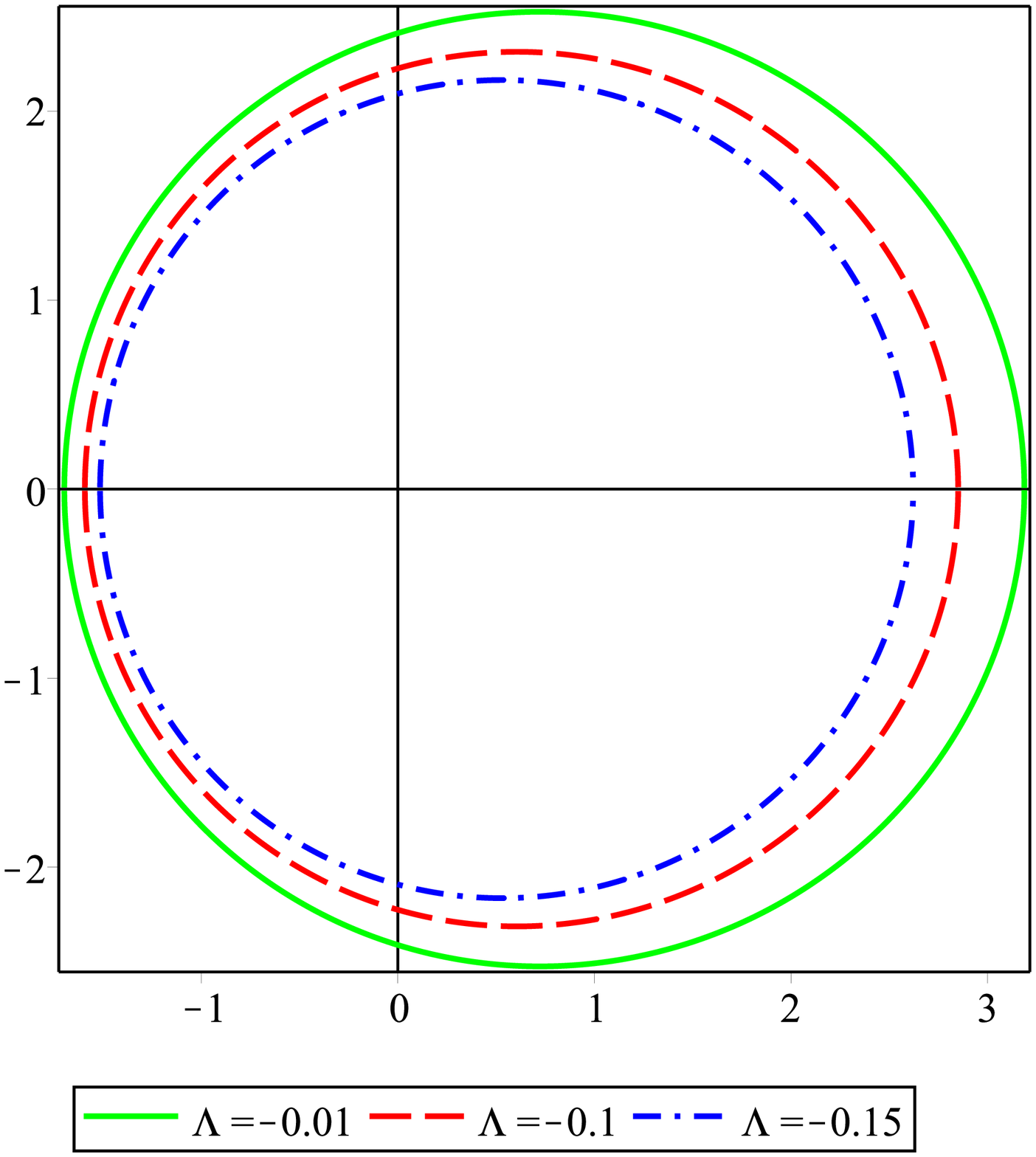}}
\caption{Shadow cast by a charged rotating BH in dRGT massive
gravity with $M=1$ and different values of $
\theta_{0} $, $ a $, $Q $,  $\gamma$,  $\varepsilon$ and $\Lambda$.}
\label{Figshnew}
\end{figure}
In the present paper, we shall use the definition adopted in Ref.
\cite{Feng1g}.  According to Eq. (\ref{xy}), the shape of
the shadow depends on the observer's viewing angle $\theta_0$. For
north pole $\theta_{0} =0$ (or equivalent south pole
$\theta_{0}=\pi$) the shadow remains a round disk. In this case,
the photons forming the shadow boundary satisfy
\begin{equation}
\xi(r^0_{\rm ph})=0\,,
\end{equation}
and the shadow radius is
\begin{equation}
R_{\rm sh}=\sqrt{\frac{a^2 + \eta(r^0_{\rm ph})}{1+\frac{\Lambda}{3}\left[a^{2}+\eta(r^0_{\rm ph})\right]}}.
\label{Rsh1}
\end{equation}

When $\theta_{0}\ne 0$ or $\pi$, the shadows are no longer round
but distorted. The maximum distortion occurs for $\theta_0=\pi/2$
in the equatorial plane. For this case, the vertical $y$-direction
is elongated while the horizontal $x$-direction is squeezed, but
the shadows remain convex. For $y=0$, there are two real solutions
\begin{equation}
\eta(r_{\rm ph}^\pm)=0\,,\qquad \hbox{with}\qquad r_{\rm ph}^+\ge  r_{\rm ph}^-
\end{equation}

The size of the shadow is defined as
\begin{equation}
R_{\rm sh} = \frac{1}{2} \big(x(r_{\rm ph}^{+}) - x(r_{\rm ph}^{-})\big)\,.\label{Rsh2}
\end{equation}

For general $\theta_0$, one can determine $r_{\rm ph}^\pm$ by requiring
\begin{equation}
y(r, \theta_0)\Big|_{r=r_{\rm ph}^\pm}=0\,,
\end{equation}
and the shadow size is then again formally given by (\ref{Rsh2}).

Here, we are interested in investigating the allowed region to
observe acceptable optical behavior for the rotating BH in dRGT
massive gravity. To do so, we have listed several values of the
event horizon, photon sphere radius, and shadow radii in table
\ref{table31}. As we see, similar to the non-rotating case, an
acceptable optical result can be obtained just for limited regions
of the electric charge, rotation parameter, cosmological constant,
and massive parameters $\gamma$ and $\varepsilon$. As a
significant point related to the effects of $Q$, $a$, and
$\varepsilon$, we should note that as these parameters increase,
some constraints are imposed on them due to the imaginary event
horizon. Regarding the influence of the cosmological constant and
the parameter $\gamma$ on the admissible region, our findings show
that increasing these two parameters leads to a non-physical
result $\frac{r_{sh}}{r_{ph}}<1$. This table also helps us find
the impact of the black hole parameters on the size of the event
horizon and photon sphere. As we see, all parameters except the
cosmological constant have a decreasing effect on the event
horizon and photon sphere radius. Although the cosmological
constant has a decreasing contribution to the event horizon
similar to other parameters, its contribution to the photon sphere
radius is opposite.

In Fig. \ref{Figshnew}, different shapes of the shadow are
obtained by plotting $ y $ against $ x $ using Eq. (\ref{xy}).
Figs. \ref{Figshnew}(a) and \ref{Figshnew}(b) display the effects
of angle $\theta_0$ and rotation parameter $ a $ on the shadow
shape. It is evident that  deformation in shapes of the shadow
gets more significant with the increasing $\theta_0$ and $ a $.
Regarding the impacts of electric charge and parameters $ \gamma $
and $ \varepsilon $ on the radius of  shadow, we see that the
shadow size shrinks with increasing of electric charge (see Fig.
\ref{Figshnew}(c)) and  massive parameters $ \gamma $ and $
\varepsilon $ (see Figs. \ref{Figshnew}(d) and \ref{Figshnew}(e)).
To study the impact of the cosmological constant on the black hole
shadow, we have plotted Fig. \ref{Figshnew}(f), indicating that
increasing $ \vert \Lambda \vert $ makes the decreasing of the
shadow radius. Furthermore, we find that variation of $ Q $ has a
weaker effect on the shadow size than the massive parameters.
\begin{figure}[H]
\centering \subfloat[$ Q=\gamma =\varepsilon=0.2$ and $
\Lambda=-0.01 $]{
        \includegraphics[width=0.31\textwidth]{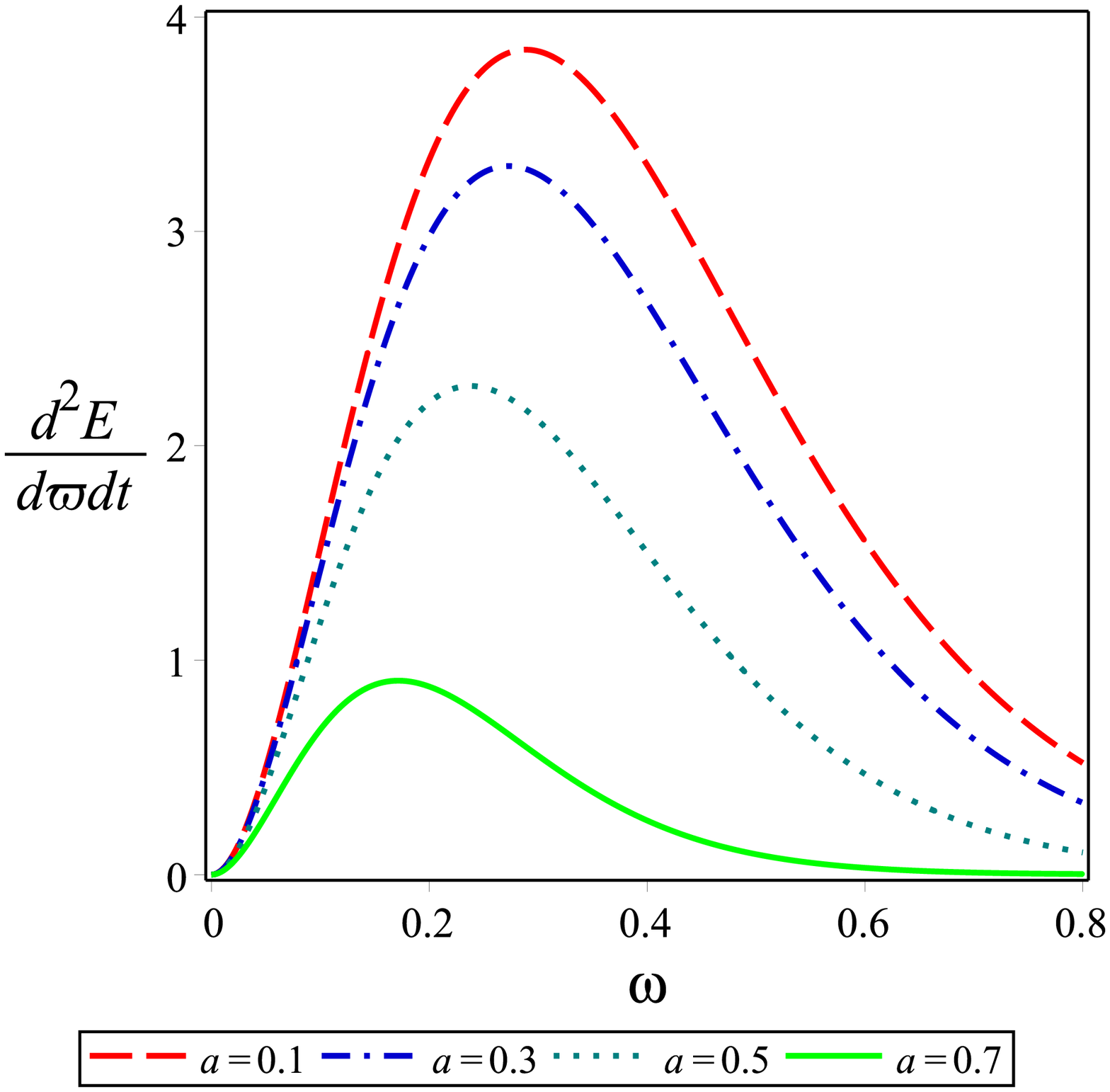}}
\subfloat[$ a=0.5$, $\gamma =\varepsilon=0.2$ and $ \Lambda=-0.01 $]{
        \includegraphics[width=0.31\textwidth]{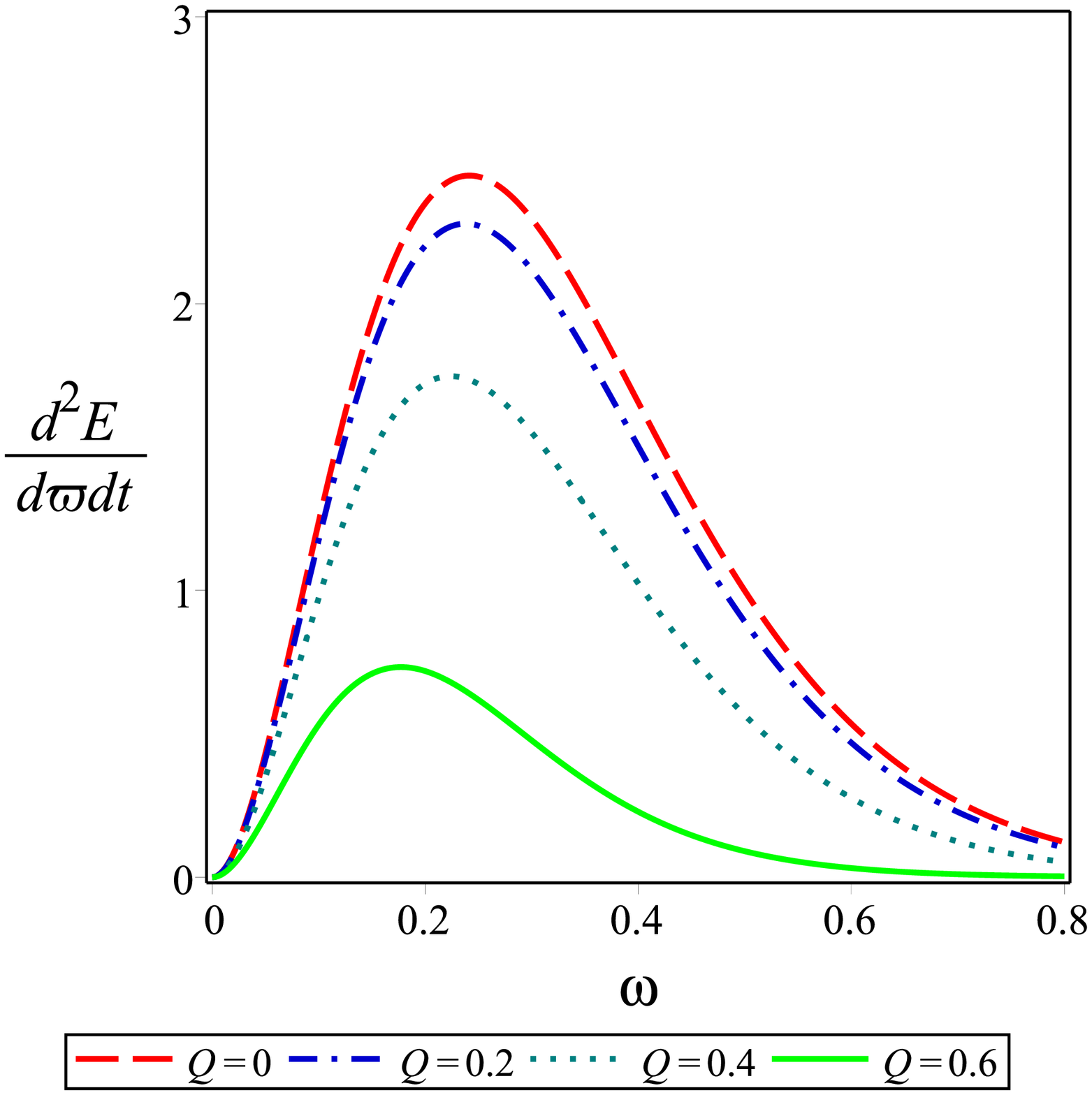}}
\subfloat[$ a=0.5$, $ Q=\varepsilon=0.2$ and $ \Lambda=-0.01 $]{
        \includegraphics[width=0.305\textwidth]{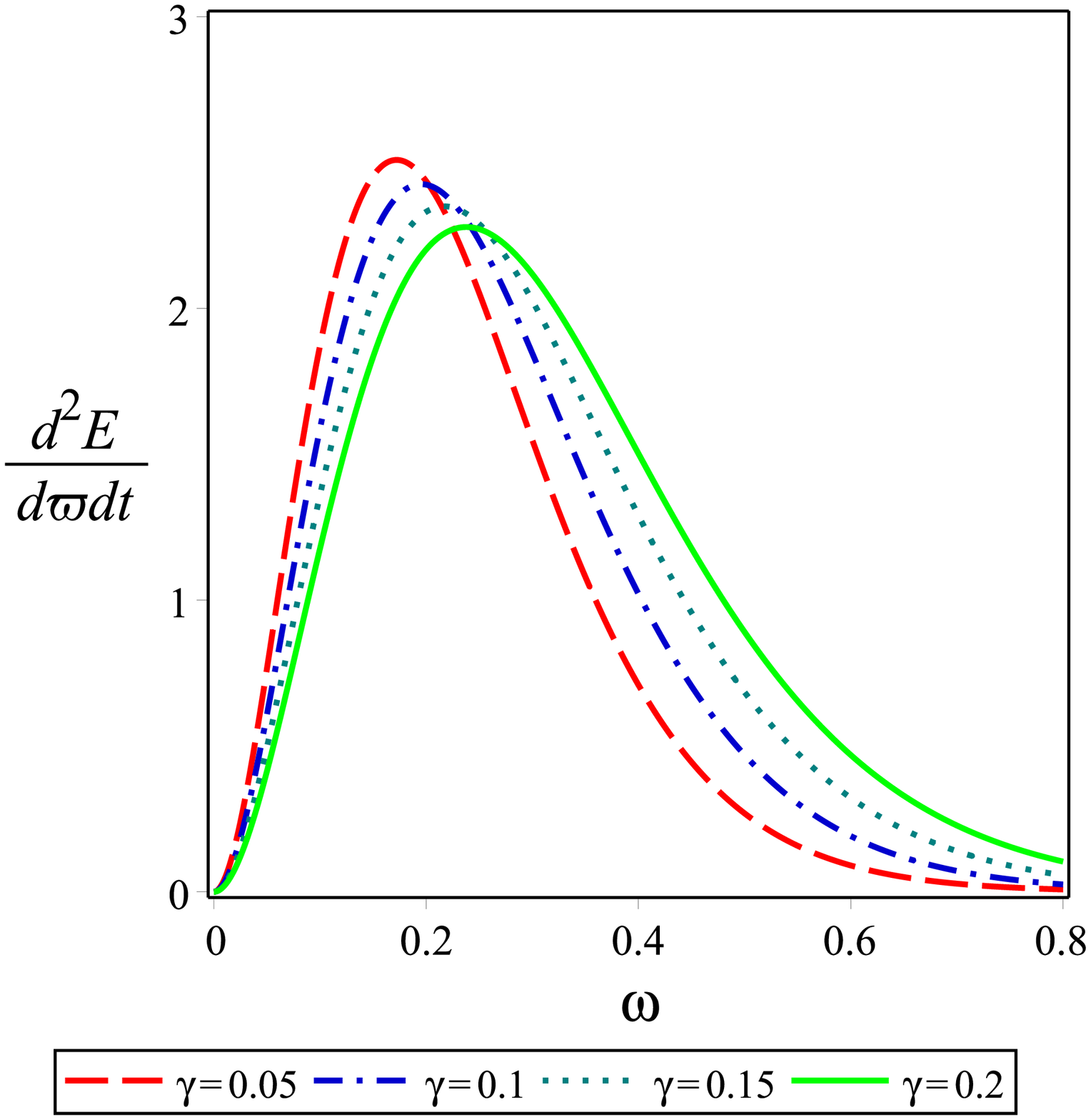}}\newline
\subfloat[ $ Q=\gamma=0.2 $, $ a=0.5$ and $ \Lambda=-0.01 $]{
        \includegraphics[width=0.31\textwidth]{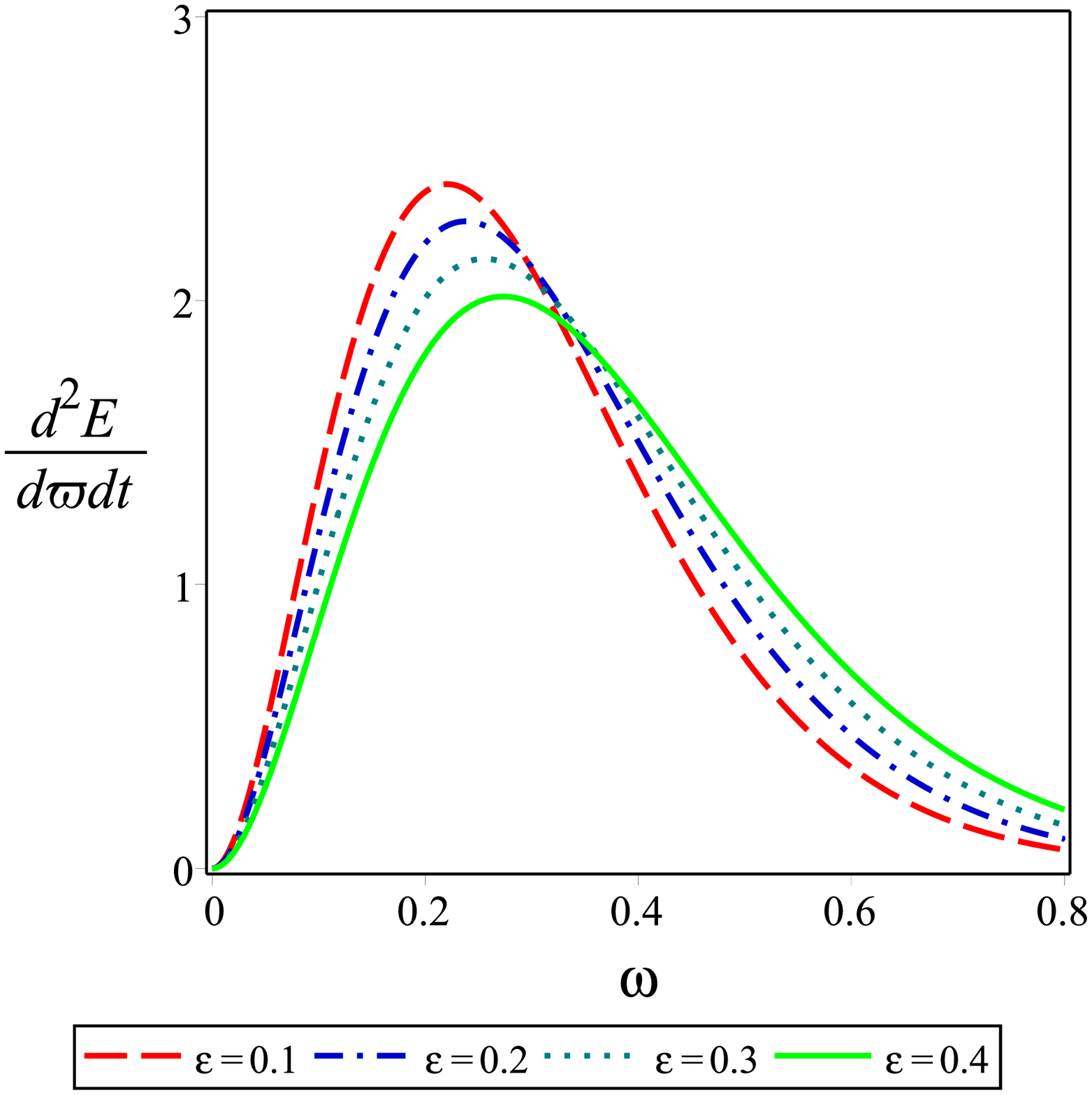}}
\subfloat[ $ Q= \gamma =\varepsilon=0.2$ and $a=0.5$]{
        \includegraphics[width=0.31\textwidth]{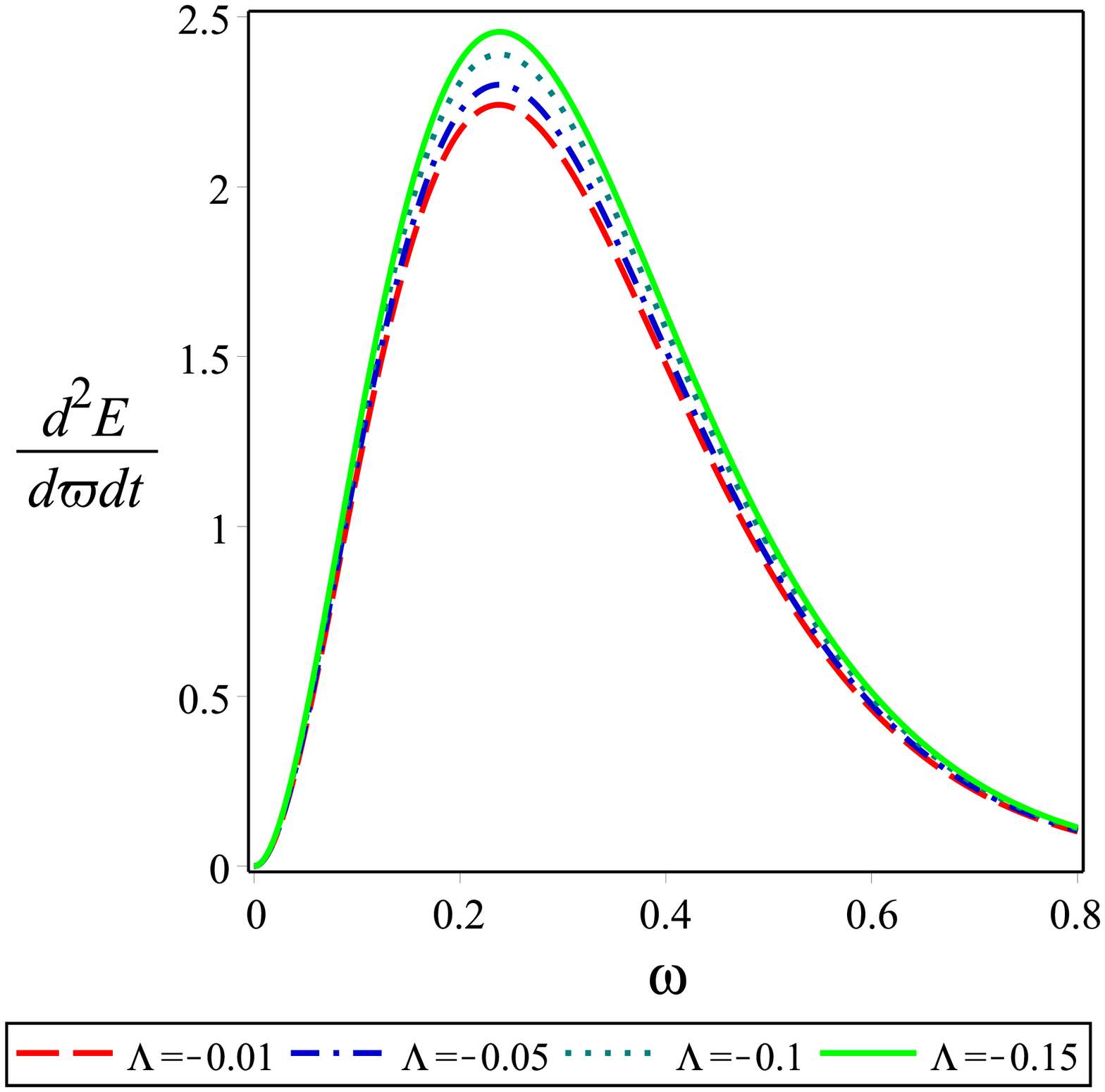}}\newline
\caption{Evolution of the BH energy emission rate with the
frequency $  \omega$ for $M=1$ and different
values of $ a $, $Q $,  $\gamma$, $\varepsilon$ and $ \Lambda $. }
\label{FigErnew}
\end{figure}

\subsection{Energy emission rate}\label{Sec. IIIB}

As it was already mentioned, for an observer located at infinity,
the BH shadow corresponds to the high energy absorption
cross-section of the BH, which oscillates around a constant
limiting value $ \sigma_{lim}\approx \pi R_{\text{sh}}^{2} $. In
this subsection, we intend to extend this result to the rotating
BH considered in this work. From the geometry of the shadow, the
observable $R_{\text{sh}}  $ which designates approximately the
size of the shadow, can be expressed in the form
\begin{equation}
R_{\text{sh}}=\frac{(x_{t}-x_{r})^{2}+y_{t}^{2}}{2|x_{t}-x_{r}|}.
\end{equation}

Equation (\ref{Eqemission}) can be used to obtain the energy
emission rate for a rotating BH as well. The Hawking temperature
of  the charged rotating BH in dRGT massive gravity at event
horizon $r_{+}$ becomes
\begin{equation}
T=\frac{\Lambda r_{e}^{4}+2\gamma r_{e}^{3}+\varepsilon r_{e}^{2}-Q^{2}-a^{2}+r_{e}^{2}}{4\pi r_{e}(r_{e}^{2}+a^{2})}.
\end{equation}

\begin{figure}[H]
\centering \subfloat[$ \gamma=0.02 $, $ \varepsilon=0.05$ and
$\Lambda=-0.01$]{
        \includegraphics[width=0.315\textwidth]{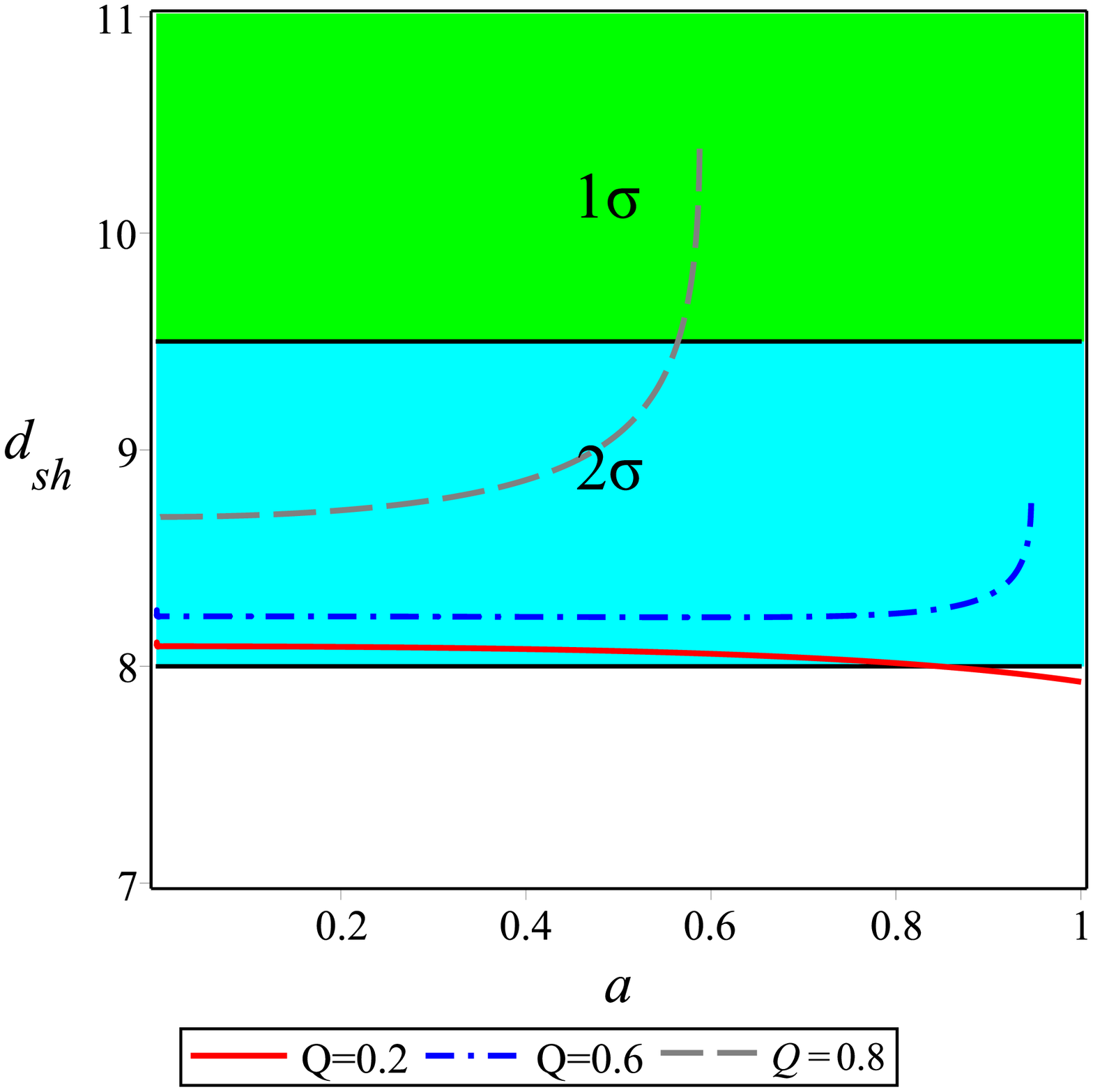}}
\subfloat[$ Q=0.8 $, $ \varepsilon=0.05$ and $\Lambda=-0.01$]{
        \includegraphics[width=0.31\textwidth]{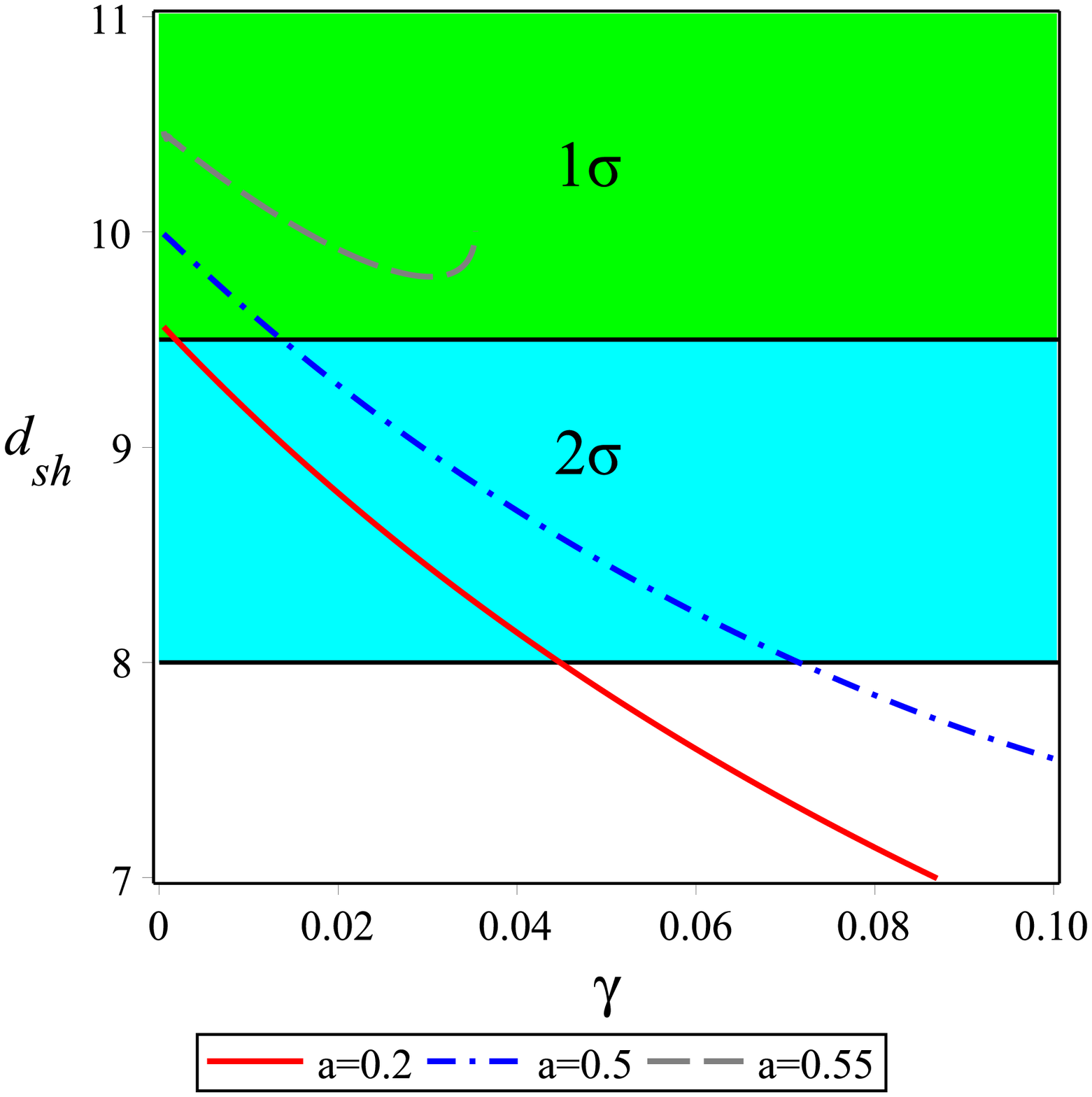}}\newline
\subfloat[$ Q=0.8 $, $ \gamma=0.02$ and $\Lambda=-0.01$]{
        \includegraphics[width=0.325\textwidth]{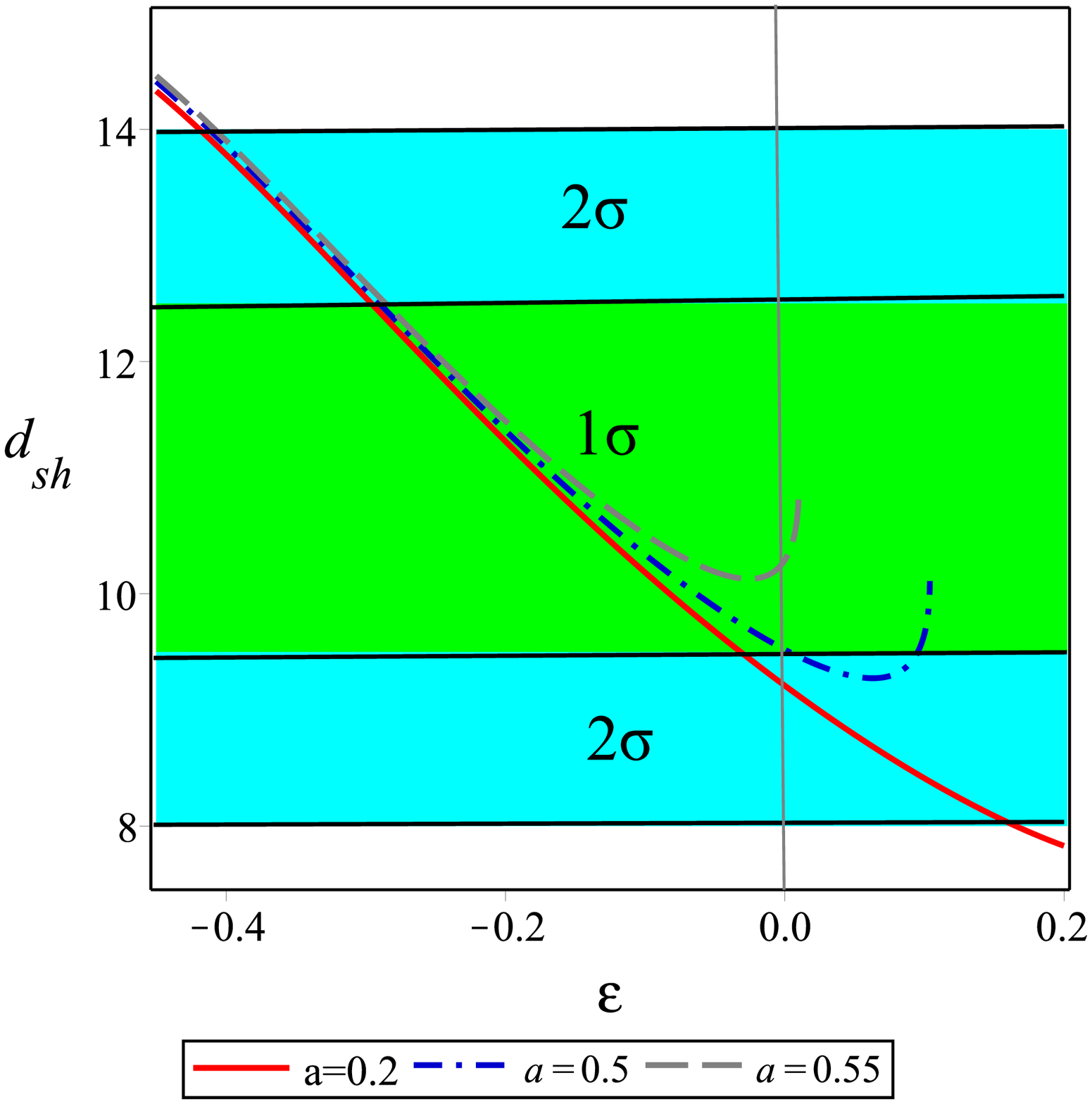}}
\subfloat[$ Q=0.8 $, $ \varepsilon=0.05$ and $\gamma=0.02$]{
        \includegraphics[width=0.31\textwidth]{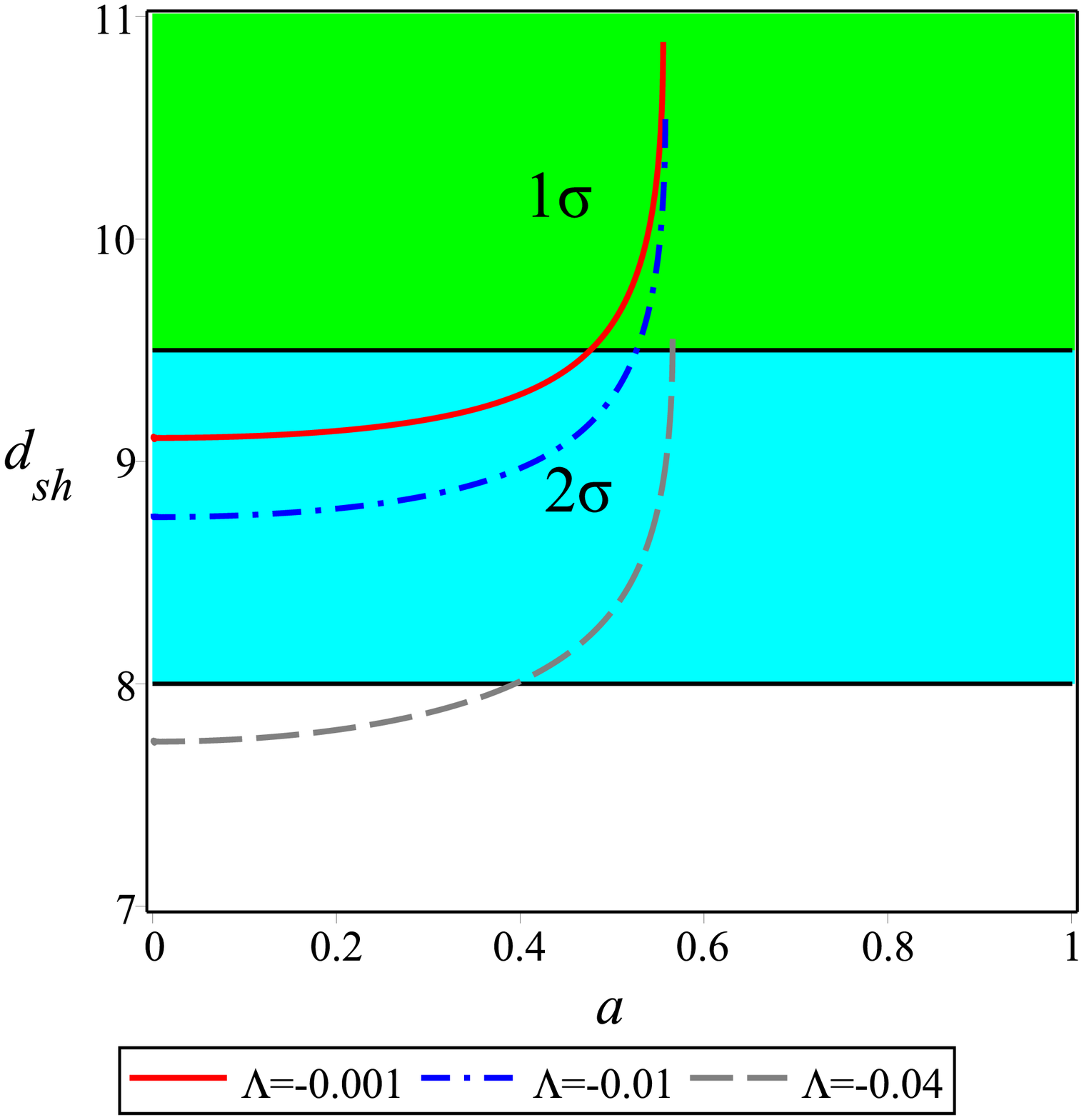}}\newline
\caption{\textbf{Left-up graph:} The predicted diameter for a
charged rotating BH in dRGT massive gravity as a function of the
rotation parameter for fixed $\gamma$, $ \varepsilon $, $ \Lambda
$, and for several values of the electric charge. \textbf{Right-up
graph:} $ d_{sh} $ as a function of the parameter $ \gamma $, for
fixed $ Q$, $ \varepsilon $, $ \Lambda $, and  different values of
the rotation parameter. \textbf{Left-down graph:} $ d_{sh} $ as a
function of the parameter $ \varepsilon $, for fixed $ Q $, $
\gamma $, $ \Lambda $, and different values of the rotation
parameter. \textbf{Right-down graph:} $ d_{sh} $ as a function of
the rotation parameter, for fixed $ Q $, $ \varepsilon $, $ \gamma
$, and  different values of the cosmological constant. The green
shaded region gives the $ 1\sigma $ confidence region for $ d_{sh}
$, whereas the cyan shaded region gives the $ 2\sigma $ confidence
region. The inclination angle is $ \theta_{0}=0^{\circ}$. }
\label{FEHTa}
\end{figure}

In Fig. \ref{FigErnew}, we illustrate the behavior of the energy
emission rate of the corresponding rotating BH against the photon
frequency $ \omega $ for various values of BH parameters. From
Figs. \ref{FigErnew}(a) and \ref{FigErnew}(b), the emission rate
peak decreases  with an increase in $a$ and $ Q $, and shifts to a
lower value of $ \omega $ with the increase of these two
parameters. This reveals that the evaporation process would be
slow for a fast-rotating BH or a BH located in a powerful electric
field. Figures. \ref{FigErnew}(c) and \ref{FigErnew}(d) display
the effective role of massive parameters $\gamma $ and $
\varepsilon $ on this optical quantity. Based on these figures,
although these parameters have a decreasing effect on the emission
rate like the electric charge and rotation parameter, the peak
of the energy emission rate shifts to a higher value of $ \omega $
with the increase of $\gamma $ and $ \varepsilon $. Taking a look
at Fig. \ref{FigErnew}(e), one can find that the effect of the
cosmological constant on the emission rate is opposite to that of
the electric charge and rotation parameter. In other words, as $
\vert \Lambda \vert $ increases, the emission of particles around
the BH increases. From what we mentioned before, one can say that
the BH will have a longer lifetime if it is located in a powerful
electric field and low curvature background or if it rotates
faster. Comparing Fig. \ref{FigErnew} to Fig. \ref{Fig5}, one can
find that although the massive parameters have an increasing
effect on the energy emission rate for static BHs in  dRGT massive
gravity, their contribution is to decrease the energy emission
rate for rotating ones. It shows that the effect of these two
parameters changes in the presence of the rotation parameter.
Also, by comparing these two figures, one can notice that a static
black hole in dRGT massive gravity has a longer lifetime if it is
located in a high curvature background, whereas its lifetime
becomes short in such a background by adding the rotation
parameter to it.

\section{Comparison with the EHT data}\label{Sec. IV}

The EHT Collaboration has recently observed the shadow of the M87*
BH residing at the center of nearby galaxy M87* using the very
large baseline interferometry technique. In this section, we
intend to compare the resulting shadow of the BHs with
observational data to find the allowed regions of the model
parameters for which the obtained shadow is consistent with the
data. Since supermassive BHs are rotating, we study the rotating
version of BHs in dRGT massive gravity to obtain more careful
results. From Fig. \ref{Figshnew}, we deduce that the size of the
resulting shadow is sensitive on the BH parameters, and hence the
image of M87* can impose bounds on them. Here, we use the EHT
constraints on the diameter of shadow to constrain the model's
parameters to be consistent with the M87* image within $ 1\sigma $
and $ 2\sigma $ confidence.

According to the obtained results by EHT collaboration
\cite{Akiyama}, the angular size of the shadow, the mass and the
distance to M87* one respectively has the values
\begin{eqnarray}
\delta &=& (42 \pm 3) \mu as,\\
M &=& (6.5 \pm 0.9) \times 10^{9} M_{\odot},\\
D &=& 16.8^{+0.8}_{-0.7} Mpc,
\label{EqdM87a}
\end{eqnarray}
where  $M_{\odot}$ is the Sun mass. These numbers imply that the
diameter of the shadow in units of mass should be \cite{Bambi1dx}
\begin{equation}
d_{M87^{*}}\equiv \frac{D\delta}{M}\approx 11.0 \pm 1.5.
\label{EqdM87b}
\end{equation}

This combination can be used to confront the theoretically
predicted shadows. As we see from Eq. (\ref{EqdM87b}),  within
$1\sigma $ uncertainty $ 9.5\lesssim d_{M87^*} \lesssim 12.5$
whereas within $2\sigma $ uncertainty $ 8.0 \lesssim d_{M87^*}
\lesssim 14.0$. In order to confront the rotating solution in dRGT
massive gravity with the above observational number, we need to
calculate the diameter of the predicted shadow.

Taking into account the allowed values of $d_{M87^*}$, we are in a
position to compare the resulting shadow of charged rotating BHs
in dRGT massive gravity with EHT data. We conduct our
investigation for the inclination angle $ \theta_{0}=0^{\circ}$,
$90^{\circ}$ and $17^{\circ} $. It is worth pointing out that in
agreement with EHT, it is better to consider the value $
\theta_{0}=17^{\circ}$ for the inclination angle.  In Fig.
\ref{FEHTa}, we plot the diameter of the resulting shadow as a
function of the rotation parameter $ a $ and massive parameters
$\varepsilon $ and $ \gamma $ for the inclination angle $
\theta_{0}=0^{\circ}$ together with $ 1\sigma $ and $ 2\sigma $
confidence intervals.

In Fig. \ref{FEHTa}(a), we see that for a more slowly rotating BH
located in a weak electric field, the resulting shadow is
consistent with observations within $ 2\sigma $-error, whereas for
a fast-rotating BH in such a situation, the resulting shadow is
not in agreement with EHT data (see the red solid curve of this
figure).  For intermediate values of the electric charge, $d_{sh}$
is located in $ 2\sigma $ confidence region for all values of the
rotation parameter (see the blue dot-dash curve in Fig.
\ref{FEHTa}(a)). For large electric charge, $d_{sh}$ becomes
consistent with EHT data within $ 1\sigma $-error only for
intermediate values of the rotation parameter, otherwise  it is in
agreement with observations in $ 2\sigma $ confidence region (see
the gray dash curve in Fig. \ref{FEHTa}(a)). As a result,  one can
say that only in the presence of a powerful electric field, the
shadow of such black holes is consistent with observational data
within $ 1\sigma$-error. Fig. \ref{FEHTa}(b) displays the behavior
of $d_{sh}$ with respect to parameter $ \gamma $ for different
values of the rotation parameter. According to our analysis, the
resulting shadow becomes consistent with the detection of the
Event Horizon Telescope in the region $ 0<\gamma <0.0445 $ for
slowly rotating BHs (see the red curve in Fig. \ref{FEHTa}(b)).
For intermediate values of the rotation parameter, a compatible
result will obtain for $\gamma<0.072$ (see the blue dot-dash curve
in Fig. \ref{FEHTa}(b)). In the region of $ 0.55<a<0.59 $, the
black hole has a shadow in agreement with observational data
within $ 1\sigma $-error (see  the gray dash curve in Fig.
\ref{FEHTa}(b)). As we see from Fig. \ref{FEHTa}(b), for too small
values of the parameter $ \gamma $, the resulting shadow of fast
(slowly) rotating BHs is in agreement with observational data
within $ 1\sigma $-error ($ 2\sigma $-error). In Fig.
\ref{FEHTa}(c), we study the allowed region of the parameter $
\varepsilon $ which is in agreement with EHT detection.  Our
investigation shows that for fast (slowly) rotating BHs the
allowed region of this parameter is $ -0.42<\varepsilon <0.01 $ ($
-0.42<\varepsilon <0.169 $).  As we see, in order to have
consistent results with EHT data, it is better to consider
negative $ \varepsilon $ values. From this figure, one can also
find that in the region $ -0.3<\varepsilon <-0.03 $, a  rotating
BH is in $ 1\sigma $ confidence region all the time. Fig.
\ref{FEHTa}(e) displays the effective role of the cosmological
constant on the allowed region. It can be seen that the black hole
should be considered in a very low curvature background in order
for the results to be consistent with the EHT data. In a high
curvature background, only fast-rotating BHs have a shadow in
agreement with observational data within the $ 2\sigma $-error
(see the gray dash curve in Fig. \ref{FEHTa}(e)).

\begin{table*}[htb!]
\centering \caption{The shadow diameter for the variation of the
rotation parameter,  the electric charge,  the parameter $\gamma$
and $ \varepsilon $ for $M=1$ and $ \theta_{0}=\pi /2 $.}
\label{table22}
\begin{tabular}{c c c c c c}
 \footnotesize $a$  \hspace{0.3cm} & \hspace{0.3cm}$0.1$ \hspace{0.3cm} &
\hspace{0.3cm} $0.3$\hspace{0.3cm} & \hspace{0.3cm} $0.5$\hspace{0.3cm} & \hspace{0.3cm}$0.8$\hspace{0.3cm} & \hspace{0.3cm} \\ \hline\hline
$ d_{sh} (Q=0.2 $, $\gamma=0.02$, $\varepsilon=0.05 $  and $ \Lambda=-0.01 $)  & $ 8.6012 $ & $8.5443$ & $8.4205$&$8.0187$ \\\hline
\\
\footnotesize $Q$  \hspace{0.3cm} & \hspace{0.3cm}$0.1$ \hspace{0.3cm} &
\hspace{0.3cm} $0.3$\hspace{0.3cm} & \hspace{0.3cm} $0.5$\hspace{0.3cm} & \hspace{0.3cm}$0.7$\hspace{0.3cm} & \hspace{0.3cm} \\ \hline\hline
$ d_{sh} (a=0.1 $, $\gamma=0.02$, $\varepsilon=0.05 $  and $ \Lambda=-0.01 $)  & $8.6424 $ & $8.5310$ & $8.2939$&$7.8967$ \\\hline
\\
\footnotesize $\gamma$  \hspace{0.3cm} & \hspace{0.3cm}$0.01$ \hspace{0.3cm} &
\hspace{0.3cm} $0.02$\hspace{0.3cm} & \hspace{0.3cm} $0.04$\hspace{0.3cm} & \hspace{0.3cm}$0.05$\hspace{0.3cm} & \hspace{0.3cm} \\ \hline\hline
$ d_{sh} (a=0.1 $, $Q=0.2$, $\varepsilon=0.05 $  and $ \Lambda=-0.01 $)  & $8.9038 $ & $8.6012$ & $8.0693$&$7.8339$ \\\hline
\\
 \footnotesize $\varepsilon$  \hspace{0.3cm} & \hspace{0.3cm}$-0.25$ \hspace{0.3cm} &
\hspace{0.3cm} $-0.1$\hspace{0.3cm} & \hspace{0.3cm} $0.05$\hspace{0.3cm} & \hspace{0.3cm}$0.1$\hspace{0.3cm} & \hspace{0.3cm}  \\ \hline\hline
$ d_{sh} (a=0.1 $, $Q=0.2$, $\gamma=0.02 $  and $ \Lambda=-0.01 $)  & $ 12.8262 $ & $10.4187$ & $8.6012$&$8.0812$ \\\hline
\\
 \footnotesize $\Lambda$  \hspace{0.3cm} & \hspace{0.3cm}$-0.001$ \hspace{0.3cm} &
\hspace{0.3cm} $-0.01$\hspace{0.3cm} & \hspace{0.3cm} $-0.03$\hspace{0.3cm} & \hspace{0.3cm}$-0.04$\hspace{0.3cm} & \hspace{0.3cm}  \\ \hline\hline
$ d_{sh} (a=0.1 $, $Q=0.2$, $\gamma=0.02 $ and  $\varepsilon=0.05 $)  & $8.8515 $ & $8.6012$ & $8.1132$&$7.8985$ \\\hline
\end{tabular}
\end{table*}

\begin{table*}[htb!]
\centering \caption{The shadow diameter for the variation of the rotation parameter,  the electric charge,  the parameter $\gamma$  and $ \varepsilon $ for $M=1$ and $ \theta_{0}=17^{\circ} $.} \label{table32}
\begin{tabular}{c c c c c c}
 \footnotesize $a$  \hspace{0.3cm} & \hspace{0.3cm}$0.1$ \hspace{0.3cm} &
\hspace{0.3cm} $0.3$\hspace{0.3cm} & \hspace{0.3cm} $0.5$\hspace{0.3cm} & \hspace{0.3cm}$0.9$\hspace{0.3cm} & \hspace{0.3cm} \\ \hline\hline
$ d_{sh} (Q=0.2 $, $\gamma=0.02$, $\varepsilon=0.05 $ and $\Lambda=-0.01$)  & $ 8.6052 $ & $8.5816$ & $8.5298$&$8.2674$ \\\hline
\\
\footnotesize $Q$  \hspace{0.3cm} & \hspace{0.3cm}$0.1$ \hspace{0.3cm} &
\hspace{0.3cm} $0.3$\hspace{0.3cm} & \hspace{0.3cm} $0.5$\hspace{0.3cm} & \hspace{0.3cm}$0.7$\hspace{0.3cm} & \hspace{0.3cm} \\ \hline\hline
$ d_{sh} (a=0.1 $, $\gamma=0.02$, $\varepsilon=0.05 $  and $ \Lambda=-0.01 $)  & $ 8.6464 $ & $8.5352$ & $8.2685$&$7.8969$ \\\hline
\\
\footnotesize $\gamma$  \hspace{0.3cm} & \hspace{0.3cm}$0.01$ \hspace{0.3cm} &
\hspace{0.3cm} $0.02$\hspace{0.3cm} & \hspace{0.3cm} $0.04$\hspace{0.3cm} & \hspace{0.3cm}$0.05$\hspace{0.3cm} & \hspace{0.3cm} \\ \hline\hline
$ d_{sh} (a=0.1 $, $Q=0.2$, $\varepsilon=0.05 $ and $\Lambda=-0.01$)  & $ 8.9072 $ & $8.6052$ & $8.0743$&$7.8393$ \\\hline
\\
 \footnotesize $\varepsilon$  \hspace{0.3cm} & \hspace{0.3cm}$-0.25$ \hspace{0.3cm} &
\hspace{0.3cm} $-0.1$\hspace{0.3cm} & \hspace{0.3cm} $0.05$\hspace{0.3cm} & \hspace{0.3cm}$0.1$\hspace{0.3cm} & \hspace{0.3cm}  \\ \hline\hline
$ d_{sh} (a=0.5 $, $Q=0.2$, $\gamma=0.02 $ and and $\Lambda=-0.01$)  & $ 12.8308 $ & $10.4597$ & $8.6052$&$8.1042$ \\\hline
\\
 \footnotesize $\Lambda$  \hspace{0.3cm} & \hspace{0.3cm}$-0.001$ \hspace{0.3cm} &
\hspace{0.3cm} $-0.01$\hspace{0.3cm} & \hspace{0.3cm} $-0.03$\hspace{0.3cm} & \hspace{0.3cm}$-0.04$\hspace{0.3cm} & \hspace{0.3cm}  \\ \hline\hline
$ d_{sh} (a=0.1 $, $Q=0.2$, $\gamma=0.02 $ and  $\varepsilon=0.05 $)  & $8.8541 $ & $8.6052$ & $8.1197$&$7.9058$ \\\hline
\end{tabular}
\end{table*}

We continue our investigation for the inclination angle $
\theta_{0}=90^{\circ}$ and $17^{\circ} $. Since analytical method
is not possible for these cases, we employ numerical method. The
diameter of shadow  for some parameters is listed in table
\ref{table22} ($ \theta_{0}=90^{\circ}$) and table \ref{table32}
($ \theta_{0}=17^{\circ}$), thereby we understand the behavior of
$d_{sh}$ under variation of the rotation parameter, electric
charge, cosmological constant, parameters  $ \gamma $ and $
\varepsilon $. In these two tables, we examine the allowed regions
of the BH parameters within $ 1\sigma $/$ 2\sigma $ -error. For
the inclination angle $ \theta_{0}=90^{\circ}$, the resulting
shadow of the charged rotating BH is in agreement with
observations for values of $ 0<a<0.8 $.  Whereas for angle $
\theta_{0}=17^{\circ}$, the allowed region of rotation parameter
is $ 0<a<0.9 $. Our findings show that for values of $
-0.203<\varepsilon<-0.03 $, the resulting shadow is in agreement
with EHT data within $ 1\sigma $-error, whereas for values of $
-031<\varepsilon<-0.203 $  or $ -0.03<\varepsilon<0.1 $  the
resulting shadow is consistent with EHT data within $ 2\sigma
$-error.

\section{Conclusion}\label{Sec. V}

Various toy models in gravitating systems and their
properties are interesting from the mathematical point of view.
However, in order to veto, confirm or restrict a theoretical model
from the physical viewpoint, we have to use observational evidence
and compare our theoretical results with the real data.

In this paper, we have considered charged AdS BHs in dRGT massive
gravity and studied their optical features such as the photon
sphere radius, shadow size, and energy emission rate in detail.
Studying the radius of photon sphere and shadow showed that some
constraints should be imposed on the range of parameters of the
model under consideration to obtain acceptable optical results. By
investigation of the energy emission rate, we have found that the
parameters of dRGT massive gravity have an increasing contribution
on the emission rate, namely, the emission of particles around the
BH increases by increasing these parameters. Regarding the role of
electric charge and cosmological constant, we have shown that they
have a decreasing effect on the emission rate, revealing the fact
that such BHs will have a longer lifetime in a stronger electric
field or a high curvature background.

As the next step, we generalized the static spherically symmetric
charged BHs in dRGT massive gravity to the corresponding rotating
solutions via the Newman-Janis algorithm. We have studied the
shadow of these BHs and found that the rotation parameter causes
deformations to both the size and shape of the BH shadow, which is
in agreement with the so-called black hole image. We also noticed
that the shape of the shadow changes depending on the inclination
angle $ \theta_{0}$ of the observer at the asymptotic infinity.
For the case of $ \theta_{0}=0$, the shadow was a round disk
regardless of the angular momentum. The shape is skewed with the
increase in $ \theta_{0}$ and became most distorted at the
equatorial plane ($ \theta_{0}=\pi/2$). Examining the effect of
electric charge, cosmological constant,  $\gamma$ and $\varepsilon$ on the radius of
shadow, we have seen that every four parameters have a decreasing
contribution on the shadow size. We continued by investigating the
energy emission rate for this rotating case and observed that the
effects of rotation parameter, electric charge, $ \gamma $ and $\varepsilon$
decrease the emission rate, whereas the effect of $\Lambda$ is
opposed to them.

Finally, we compared the resulting shadow of the charged rotating
BH in dRGT massive gravity to the data of EHT collaboration in
order to find the allowed values of model parameters. We conducted
our investigation for three different inclination angles $
\theta_{0}=0^{\circ}$, $90^{\circ}$ and $17^{\circ} $. For the
case of $ \theta_{0}=0^{\circ}$, we found that the resulting
shadow of a fast-rotating BH with a strong electric field is
consistent with observations within $ 1\sigma $-error. While a
fast-rotating BH with a weak electric field could not have a
shadow consistent with EHT data. Regarding massive parameter
$\gamma$, the results showed that just for small values of this
parameter ($\gamma <0.072$), one can observe consistent results to
EHT detection. Our analysis also indicated that the massive
parameter $\varepsilon$ has a significant role in the allowed
region such that for values of $-0.3<\varepsilon <-0.03 $, a
rotating BH is in the $ 1\sigma $ confidence region all the time.
Examining the cosmological constant effect, we found that a
rotating BH can have a consistent result with EHT data just in a
very low curvature background. We studied the diameter of shadow
for inclination angle $ \theta_{0}=90^{\circ}$ and $17^{\circ} $
and noticed that the allowed region of the rotation parameter for
the inclination angle $ \theta_{0}=90^{\circ}$, is smaller than
that of $\theta_{0}=17^{\circ}$. Nonetheless the allowed regions
related to other parameters were the same for both cases.

 Although in this work, the reference metric was taken
to be non-dynamical, there are many reasons that one would like to
employ a dynamical reference metric. From a theoretical point of
view,  promoting the reference metric to a full dynamical one is
desirable as leads to a background-independent theory which is
invariant under general coordinate transformations, without the
introduction of St\"{u}ckelberg fields \cite{bimetric}. Such
theories known as bimetric theories of gravity have renewed
interest due to their accelerating cosmological solutions
\cite{Damour:2002}. Moreover, it has been shown that having a null
apparent horizon, a number of interesting properties for the black
hole solutions are obtained by considering the massive graviton
\cite{Rosen2018kl}. So, it would be interesting to perform such an
analysis in bimetric theory or study the presence of a
time-dependent  black hole. We leave this investigation for
further work.

\begin{acknowledgements}
We would like to thank the anonymous referee for constructive
comments. SHH and KhJ thank Shiraz University Research Council.
KhJ is grateful to the Iran Science Elites Federation for the
financial support. BEP thanks University of Mazandaran.
\end{acknowledgements}

\begin{center}
\textbf{Appendix A: Degrees of Freedom} \label{A}

\end{center}
Before studying the number of degrees of freedom in the massive
theory of gravity, we briefly summarize how the counting of the
number of degrees of freedom can be performed in the ADM language
using the Hamiltonian for GR. According to ADM formulation, a
metric is given by \cite{Rham2014:a}
\begin{equation}
ds^{2}=-N^{2}dt^{2}+\gamma_{ij}(dx^{i}+N^{i}dt)(dx^{j}+N^{j}dt),
\end{equation}
with the lapse $N$, the shift $ N_{i} $ and the 3-dimensional
space metric $ \gamma_{ij} $ which are parameterized in the
following way
\begin{equation}
N=(-g^{00})^{-\frac{1}{2}},~~~~~N_{i}=g_{0i},~~~~~\gamma_{ij}=g_{ij}.
\end{equation}

The Hamiltonian density for GR is expressed as
\begin{equation}
\mathcal{H}_{GR}=N\mathcal{R}_{0}(\gamma,p)+N^{i}\mathcal{R}_{i}(\gamma,p),
\end{equation}
where $ p $ is the conjugate momentum associated with $ \gamma $.
As one can see, $ \mathcal{H}_{GR} $ depends linearly on the shift
or the lapse. So they act as Lagrange multipliers and propagate a
first-class constraint each which removes 2 phase space degrees of
freedom per constraint. The number of degrees of freedom in phase
space is obtained by the following formula
\begin{equation}
(2\times 6)-2 ~\text{lapse constraints}- 2 \times 3 ~\text{shift constraints} = 4 = 2 \times 2
\end{equation}
which corresponds to 4 degrees of freedom in phase space, or 2 independent degrees of
freedom in field space. It should be noted that $ \gamma_{ij} $ and $  p^{ij}$ carry $  6$ component each.

In massive gravity the Hamiltonian density is defined as \cite{Rham2014:a}
\begin{equation}
\mathcal{H}=N\mathcal{R}_{0}(\gamma,p)+N^{i}\mathcal{R}_{i}(\gamma,p)+m_{g}^{2}\mathcal{U}(\gamma_{ij},N^{i},N),
\end{equation}

Here the lapse and shift are non-linear in the Hamiltonian density
and they are no longer directly Lagrange multipliers (see Refs.
\cite{HassanI,Rham2014:a,Schmidt2012} for more details). So, they
do not propagate a constraint for the metric. Consequently,  for
an arbitrary potential, there are $ (2\times 6) $ degrees of
freedom in the three-dimensional metric and its momentum conjugate
and no constraint is present to reduce the phase space. Therefore,
a massive spin-2 field propagates 6  degrees of freedom in field
space. These $ 6 $ polarizations correspond to the five healthy
massive spin-2 field degrees of freedom in addition to a BD ghost
as the sixth degree of freedom. As we have seen, in GR the four
constraint equations of the theory along with four general
coordinate transformations remove four of the six propagating
modes of the metric. While in massive gravity the four constraint
equations generically remove the four non-propagating components
\cite{HassanI}. To solve this problem, a theory of massive gravity
should be constructed in such a way that one of the constraint
equations and an associated secondary constraint eliminate the
propagating ghost mode. Although the linear Fierz-Pauli theory
succeeds in eliminating the ghost in this way\footnote{At the
linear level of Fierz-Pauli massive gravity, the lapse remains
linear, so it acts as a Lagrange multiplier generating a primary
second-class constraint and eliminating the BD ghost.},  Boulware
and Deser showed that the ghost generically reappears at the
non-linear level \cite{BDghost}. In Ref. \cite{Arkani-Hamed}, the
authors showed that the ghost is related to the longitudinal mode
of the Goldstone bosons associated with the broken general
covariance. They have resolved these problems by tuning the
coefficients in an expansion of the mass term in powers of the
metric perturbation and of the Goldstone mode. Afterward, de Rham
and Gabadadze obtained such an expansion that was ghost free in
the decoupling limit \cite{dRGTI}. Later in \cite{dRGTII}, they
constructed a non-linear massive action starting from the
decoupling limit data. Using this theory it becomes possible to
study the ghost away from the decoupling limit. This was the first
successful construction of potentially ghost-free non-linear
actions of massive gravity. In \cite{HassanI}, Hassan and Rosen
considered dRGT massive gravity based on a flat reference metric
and demonstrated propagation of a massive spin-2 field with 5
degrees of freedom. Later in \cite{Schmidt2012}, they presented
nonlinear massive gravity based on a general reference metric and
showed that this model also contains exactly five degrees of
freedom. Regarding how to remove the sixth mode, we should point
out that in the ADM language, the BD ghost is a consequence of the
absence of the Hamiltonian constraint \cite{Schmidt2012}. As we
have already mentioned, in a generic non-linear extension of
massive gravity, the mass term depends non-linearly on the
$N_{\mu}$ and no constraint impose on the $ \gamma_{ij} $ and
their conjugate momenta $ p^{ij} $. A ghost-free theory of massive
gravity must include a single constraint on $ \gamma_{ij} $ and $
p^{ij} $ along with an associated secondary constraint to
eliminate the sixth mode \cite{HassanI}. By performing the linear
change of variable $ N_{i} \rightarrow n_{i}=\frac{N_{i}}{N} $ and
substituting back into the Hamiltonian, $\mathcal {H}$ is linear
in the lapse. Thus the $ N $ equation becomes a constraint on the
$ \gamma_{ij} $ and $ p^{ij} $. Along with a secondary constraint,
this removes the ghost \cite{HassanI,Rham2014:a}. In Ref.
\cite{HassanI} was shown that when criterion 1 is satisfied, 2 is
also followed automatically.

Regarding the corresponding theory in this paper, since we
consider a non-dynamic reference metric, there are $12$ phase
space variables (components of $ \gamma_{ij} $ and $ p^{ij} $). By
redefining the fields $ N_{i} \rightarrow n_{i} $, the Hamiltonian
is linear in $ N$, indicating that Hamiltonian constraint can be
preserved here (see Ref. \cite{Vegh2013} and Refs. [37-39]
therein). So a massive spin$-2$ field propagates $5$ degrees of
freedom. Since action (\ref{action:new}) includes the Maxwell
kinetic term and using the fact that a massless vector field $
A_{i} $ in $4-$dimensions only propagates two degrees of freedom
\cite{Rham2014:a}, the theory propagates in total $7$ degrees of
freedom.

\end{document}